\documentclass{aa}
\usepackage{graphicx}
\usepackage{txfonts}
\usepackage{natbib}
\bibpunct{(}{)}{;}{a}{}{,} % to follow the A&A style

\begin{document}
   \title{Dust processing in photodissociation regions}

   \subtitle{Mid-IR emission modelling}

   \author{M. Compi\`egne\inst{1,2}
           \and
            A. Abergel\inst{1}
           \and
            L. Verstraete\inst{1}
           \and
            E. Habart\inst{1}
       }

   \offprints{M Compi\`egne email: {\bf compiegne@cita.utoronto.ca}}

   \institute{Institut d'Astrophysique Spatiale, UMR8617, CNRS, Universit\'e Paris-sud XI,
              b\^atiment 121, F-91405 Orsay Cedex, France
             \and
             Canadian Institute for Theoretical Astrophysics, University of Toronto, 
             60 St. George Street, Toronto, ON M5S 3H8, Canada
}

   \date{Received ; accepted}

 \abstract
{Mid-infrared spectroscopy of dense illuminated ridges (or photodissociation regions, PDRs) suggests dust evolution.
Such evolution must be reflected in the gas physical properties through processes like photo-electric heating
or H$_2$ formation.
}
{With Spitzer Infrared Spectrograph (IRS) and ISOCAM data, we study the mid-IR emission of closeby, well known PDRs. 
Focusing on the band and continuum dust emissions, we follow their relative contributions and analyze their variations 
in terms of abundance of dust populations.}
{In order to disentangle dust evolution and excitation effects, we use a dust emission model that we couple to radiative transfer. 
Our dust model reproduces extinction and emission of the standard interstellar 
medium that we represent with diffuse high galactic latitude clouds called Cirrus. We take the properties of dust in Cirrus as a reference 
to which we compare the dust emission from more excited regions, namely the Horsehead and the reflection nebula NGC 2023 North.}
{We show that in both regions, radiative transfer effects cannot account for 
the observed spectral variations. We interpret these variations in term of changes of the relative abundance between 
polycyclic aromatic hydrocarbons (PAHs, mid-IR band carriers) and very small grains (VSGs, mid-IR continuum carriers).
}
{We conclude that the PAH/VSG abundance ratio is 2.4 times smaller at the peak emission of the Horsehead nebula than in the Cirrus case.  
For NGC2023 North where spectral evolution is observed across the northern PDR, 
we conclude that this ratio is $\sim$\,5 times lower in the dense, cold zones of the PDR
than in its diffuse illuminated part where dust properties seem to be the same as in Cirrus.
We conclude that dust in PDRs seems to evolve from "dense" to "diffuse" properties at the small spatial scale of the dense illuminated ridge.
}
% 5 {} token are mandatory
 

   \maketitle
%
%________________________________________________________________

\section{Introduction}
Evolution of interstellar dust was first inferred from
measurements and modelling of the ultraviolet(UV)-visible extinction
\citep[]{cardelli89, fitzpatrick86, fitzpatrick88, fitzpatrick90,
  kim94a} and from the infrared (IR) colour variations in IRAS and
COBE data \citep[][]{boulanger90, abergel94, abergel96, laureijs91}.
The evolution of interstellar matter is powered by stars which have a
profound influence on the chemical and energetic balance of the
interstellar medium. Along this lifecycle, dust grains undergo
efficient processing by shocks and stellar photons which is primarily
reflected in the dust size distribution \citep[see for
  instance][]{jones97, jones2004}.  Such evolution is expected to be
particularly significant between the dense and diffuse galactic
environments.  For instance, the presence of small dust particles in
the diffuse galactic medium is necessary to account for the
near-infrared (NIR) and mid-infrared (mid-IR) observed emission
\citep[e.g.][]{Sellgren84, draine85} while in dense molecular
environment these small particles must coagulate with
the larger species \citep[e.g.][]{stepnik2003}.  Dust has a strong
impact on interstellar gas because of efficient coupling such as gas
heating by the photoelectric effect or the formation of H$_2$ on grain
surfaces. In addition, these gas-grain processes are dominated by the
contribution of small dust grains of radius $a<100$ nm
\citep[e.g.][]{habart2001, weingartner2001c, habart2004}.

\begin{figure*}
   \centering
     \includegraphics[width=0.62\textwidth]{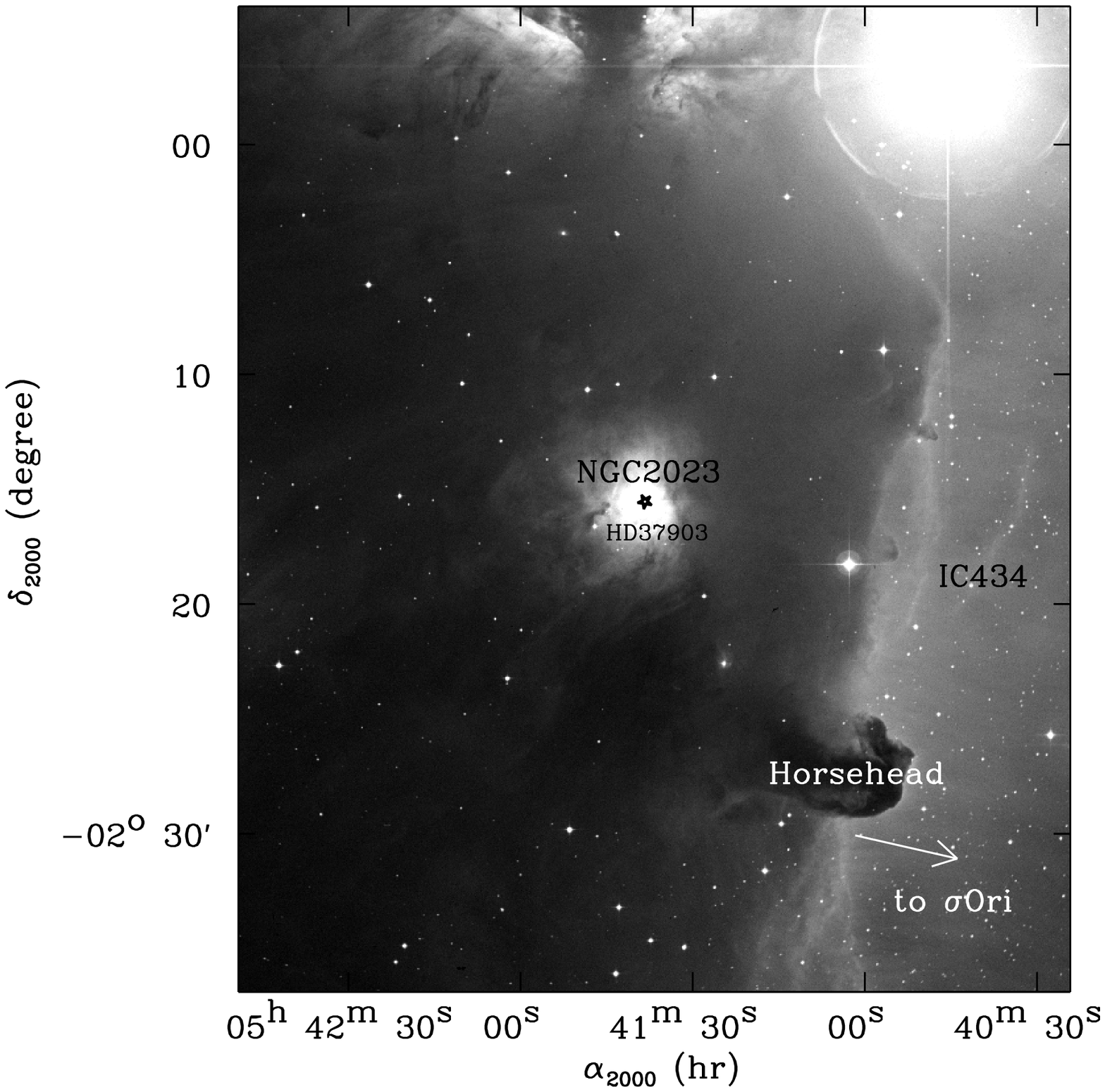}
     \includegraphics[width=0.36\textwidth]{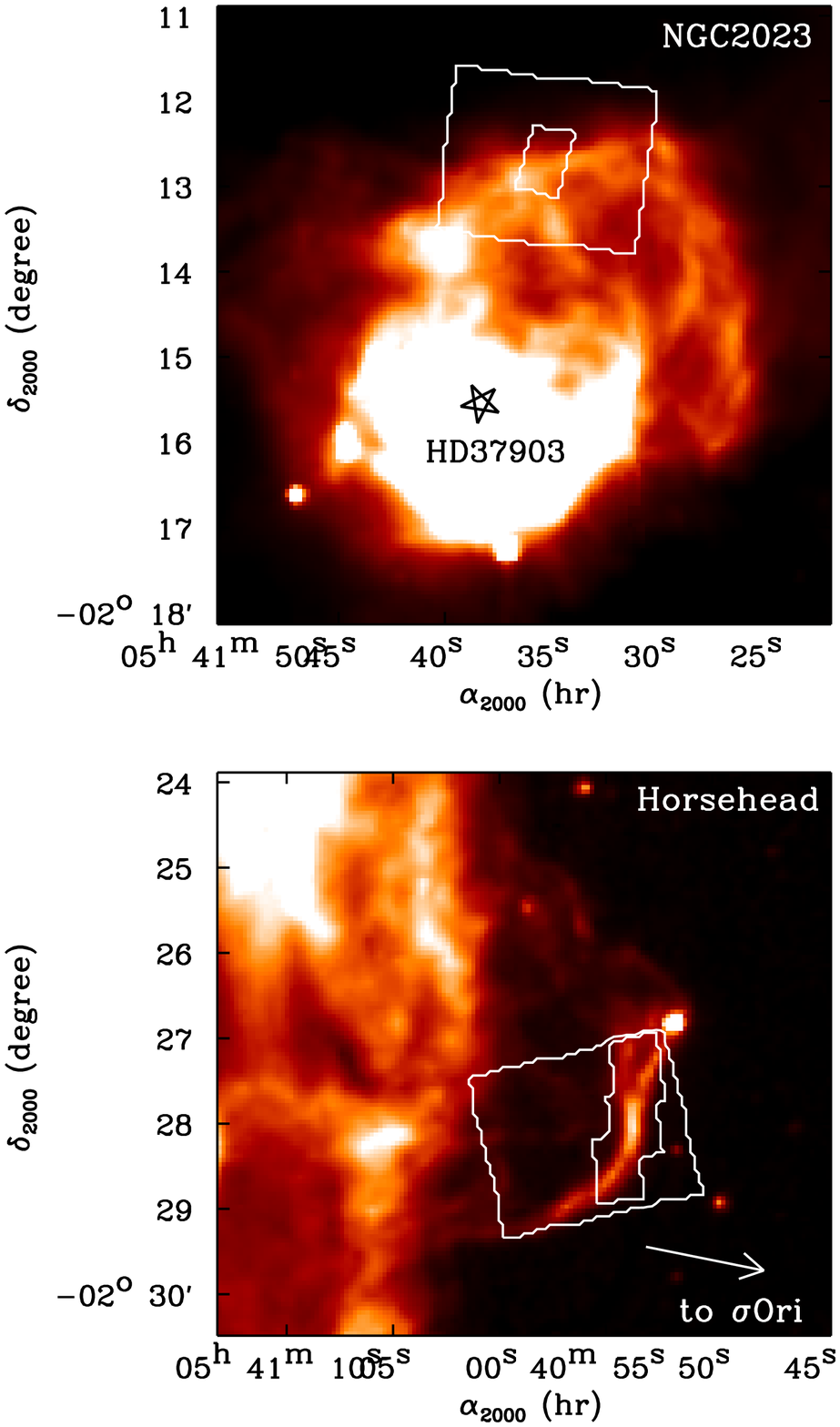}
      \caption{{\bf Left panel:} The L1630 molecular cloud 
                     as seen in the visible (photographic plate obtained
                     using the UK Schmidt Telescope and extracted from the 
                     Digitized Sky Survey produced at the Space Telescope Institute).
                     The arrow indicates the direction 
                     of the $\sigma$\,Orionis binary. 
                     {\bf Right panel:}
                     NGC2023 (top) and the Horsehead nebula (bottom) as seen 
                     by ISOCAM-LW2 at 5-8.5\,$\mu$m \citep[][]{abergel2002}.
                     The contours show areas
                     observed with Spitzer-IRS as part of the ``SPECPDR'' program \citep[][]{joblin2005}. 
                     The small and large areas delineated by the white line are those
                     observed with the SL and LL modules, respectively.}
         \label{fig:OrionB}
\end{figure*}

Located at the illuminated edges of molecular clouds,
photodissociation regions \citep[hereafter PDRs; for a review see
  e.g.][]{hollenbach97} are regions of transition between dense and
diffuse media.  In addition to the generally large density gradient
observed in such regions, there is a gradient of the UV-visible
radiation field intensity due to the proximity to a (some) young
star(s) and to the absorption of the radiation field by the dust.
Consequently, PDRs are the site of significant evolution of both the
dust and gas properties which are closely interlinked.  Dust evolution
may be a key to the understanding of the unexpectedly large amount of
warm H$_2$ in PDRs \citep[e.g.][]{allers2005, habart2008}.  PDRs are
responsible for the reprocessing of most of the UV-visible radiative
energy output from stars which is re-emitted in infrared-millimeter
wavelengths \citep[e.g.][]{hollenbach97} and consequently are
important IR emitters of normal star forming galaxies. It is thus
necessary to understand in detail the emission coming from PDRs in
order to interpret IR galaxy spectra.

Infrared Space Observatory (ISO) observations showed a systematic
diminution of the emission ratio between aromatic infrared bands (AIBs
between 3 and 17\,$\mu$m) emitted by polycyclic aromatic hydrocarbons
(PAHs) and the mid-IR continuum emission from dense illuminated ridges
\citep[e.g.][]{cesarsky2000,abergel2002}.  \citet{rapacioli2005} have
interpreted this evolution in terms of the chemical evolution of small
carbonaceous particle properties by decomposing ISOCAM/CVF spectral
cubes (5-16\,$\mu$m) with the Single Value Decomposition method.
\citet{berne2007} have extended this study by using the Blind Signal
Separation method on Spitzer IRS data.  A limitation of these methods
is that they do not take into account the radiative transfer of
stellar photons which may affect the shape of the mid-IR spectrum just
as dust properties and abundances do.  The aim of this paper is to
study the dust evolution in two well known PDRs as seen by Spitzer/IRS
and ISOCAM through the modelling of the dust mid-IR emission including
radiative transfer. Although other dust properties like the size
distribution and/or composition can change, we will interpret observed
mid-IR spectral variations in term of PAH/VSG relative abundance.  In
section \ref{sect:presentation} and \ref{sect:obs}, we present the
studied PDRs and the Spitzer data.  The section \ref{sect:result_obs}
is devoted to the observed mid-IR spectral shape in these PDRs.  In
section \ref{sect:dust_PDR_model}, we present our dust model
(\S\,\ref{sect:dust_model}) and the radiative transfer
(\S\,\ref{sect:transfer_model}) used to model dust emission in PDRs.
Sections \ref{sect:model_HH} and \ref{sect:model_2023} present results
of the modelling of the Horsehead Nebula and NGC2023 respectively. We
conclude in section\,\ref{sect:conclusion}.

\section{The selected photodissociation regions}\label{sect:presentation}

The Horsehead nebula and NGC2023 are located in the L1630 molecular cloud
as seen in the left panel of Fig.\,\ref{fig:OrionB}.  This molecular
cloud is located at a distance of $\sim$\,400\,pc \citep[from the study
of the distances to B stars in the Orion association by] []{Anthony82}
in the southern part of the Orion\,B molecular complex.

\subsection{The Horsehead nebula}

In the visible, the Horsehead nebula, also known as B33
\citep{Barnard19}, emerges from the west edge of L1630 as a dark cloud
hiding the H$\alpha$ emission of the IC434 HII region (see
left panel of 
Fig.\,\ref{fig:OrionB}).  The Horsehead nebula is a familiar object in
astronomy and has been observed many times at visible, IR and submm
wavelengths \citep[][for the 21st century only]{abergel2003, pound2003,
  teyssier2004, pety2005, hily-blant2005, habart2005,
  Ward-Thompson2006,philipp2006, Johnstone2006, goicoechea2006,
  pety2007, compiegne2007}. Both the IC434 HII region and the
Horsehead nebula PDR are mainly excited by $\sigma$\,Orionis which is an O9.5V
binary system \citep{warren77} with an effective temperature of
$\sim$\,34\,600\,K \citep{Schaerer97} and whose direction is indicated
by the arrow in Fig.\,\ref{fig:OrionB}.  
Assuming that $\sigma$\,Orionis and the Horsehead are in the
same plane perpendicular to the line of sight, the distance between
them is $\sim$\,3.5\,pc ($\sim$\,0.5\,deg) which gives
$G_0$\,$\sim$\,100
\citep[the energy density of the radiation field
 between 6 and 13.6 eV in units of the Habing field,][]{Habing68} for the
radiation field which illuminates the Horsehead nebula.

The edge of the Horsehead
nebula is delineated by an IR emission filament as seen in the right panel of
Fig.\,\ref{fig:OrionB}. \citet{abergel2002} proposed that this filament
is due to the simultaneous presence of dense material and a UV radiation
field from $\sigma$\,Orionis. The ISOCAM
5-8.5\,$\mu$m data seen on the right
panel of Fig.\,\ref{fig:OrionB} shows the mid-IR band emission which is
known to scale with the UV radiation field for $G_0$ ranging
from 1 to 10$^4$ as expected for grains small enough (PAHs) to undergo temperature spikes \citep{boulanger98b}. 
Thus, matter located deeper in the dense cloud is less excited, which explains this bright narrow IR filament. 
The Horsehead PDR is an almost plane-parallel slab seen
edge-on and is thus commonly used as a benchmark 
for comparison between models and observations \citep[e.g.][]{pety2007b}.

\subsection{NGC2023}

NGC2023 is primarily known as a reflection nebula and thus appears 
as an extended source at visible wavelengths
(left panel of Fig.\ref{fig:OrionB}).  The nebula is excited a B1.5V star (HD37903) which is embedded in the
L1630 cloud \citep[e.g.][]{gatley87} and thus illuminates the foreground dust
\citep[e.g.][]{burgh2002}
which efficiently scatters visible photons
\citep[e.g.][]{draine2003}. The position of this star, which has an
effective temperature of 23700\,K, is shown on
Fig.\ref{fig:OrionB}. Its distance has been estimated by Hipparcos
\citep[][]{perryman97} to be 470$\pm$290\,pc.  The right panel of
Fig.\ref{fig:OrionB} shows NGC2023 as seen by the LW2 channel of ISOCAM at 5-8.5\,$\mu$m
which traces mostly AIBs emitted by PAHs.

The bubble geometry of NGC2023 reflects the influence of the young star which blows up the surrounding matter 
and carves a cavity in the progenitor cloud \citep[e.g.][]{castor75, freyer2003}. This scenario is supported by the low
extinction toward HD37903 \citep[$A_{V}\sim$\,1.4,][]{burgh2002} while the total
extinction through L1630 is $A_{V}\sim$\,40 \citep[][]{harvey80}. As seen in Fig.\ref{fig:OrionB}, NGC2023 is thus 
basically a spherical cavity seen face-on.

In this work, we focus on NGC2023 North (hereafter NGC2023N)
which exhibits a strong variation of its mid-IR spectrum as shown by
\citet{abergel2002}.

\section{Observations}\label{sect:obs}

\subsection{Infrared spectrograph data}

The Horsehead Nebula and NGC2023N were observed as part of the
"SPECPDR" program \citep{joblin2005}, using the infrared spectrograph
\citep[IRS,][]{houck2004} which is a slit spectrograph on board
Spitzer \citep{werner2004}. The present observations
were performed with the {\em spectral mapping} mode and using the short-low (SL,
5.2-14.5\,$\mu$m, slit size: 57\arcsec\,$\times$\,3.6\arcsec,
R=64-128) and long-low (LL, 14-38\,$\mu$m, slit size:
168\arcsec\,$\times$\,10.6\arcsec, R=64-128) modules of the
instrument.
The integration times were 14 and 60\,s per
pointing for the second (5.2-8.7\,$\mu$m) and the first
(7.4-14.5\,$\mu$m) orders of SL, respectively, and 14 s per pointing
for both orders of LL.

Starting from BCD products, we have developed a pipeline which builds spectral cubes (two spatial
dimensions and one spectral dimension) in a homogeneous way from the
data delivered by the Spitzer Science Center (SSC). This pipeline has been presented in
\citet{compiegne2007}. 
The LL data between 35 and 38\,$\mu$m are not considered 
due to the strong decrease of sensitivity at these wavelengths.
We therefore obtain 5 to 35\,$\mu$m spectral cubes
for the observed area.
The absolute photometric uncertainty is $\sim$10\%.

The zodiacal emission Z($\lambda$) must be subtracted from these data.  
We assume that this contribution does not vary across the observed
area of the sky for a given object.  We estimate Z($\lambda$) using the
SSC background estimator\footnote{See
  http://ssc.spitzer.caltech.edu/documents/background/} which is based
on the COBE/DIRBE model \citep[][]{kelsall98}. Following 
\citet{compiegne2007}, we take
into account the zodiacal contribution in the dark observation 
(at position
RA\,=\,268$\,\fdg$96, DEC\,=\,65$\,\fdg$43 in the sky) in the BCD
level of the IRS data.

\begin{figure}
   \centering
    \includegraphics[width=0.48\textwidth]{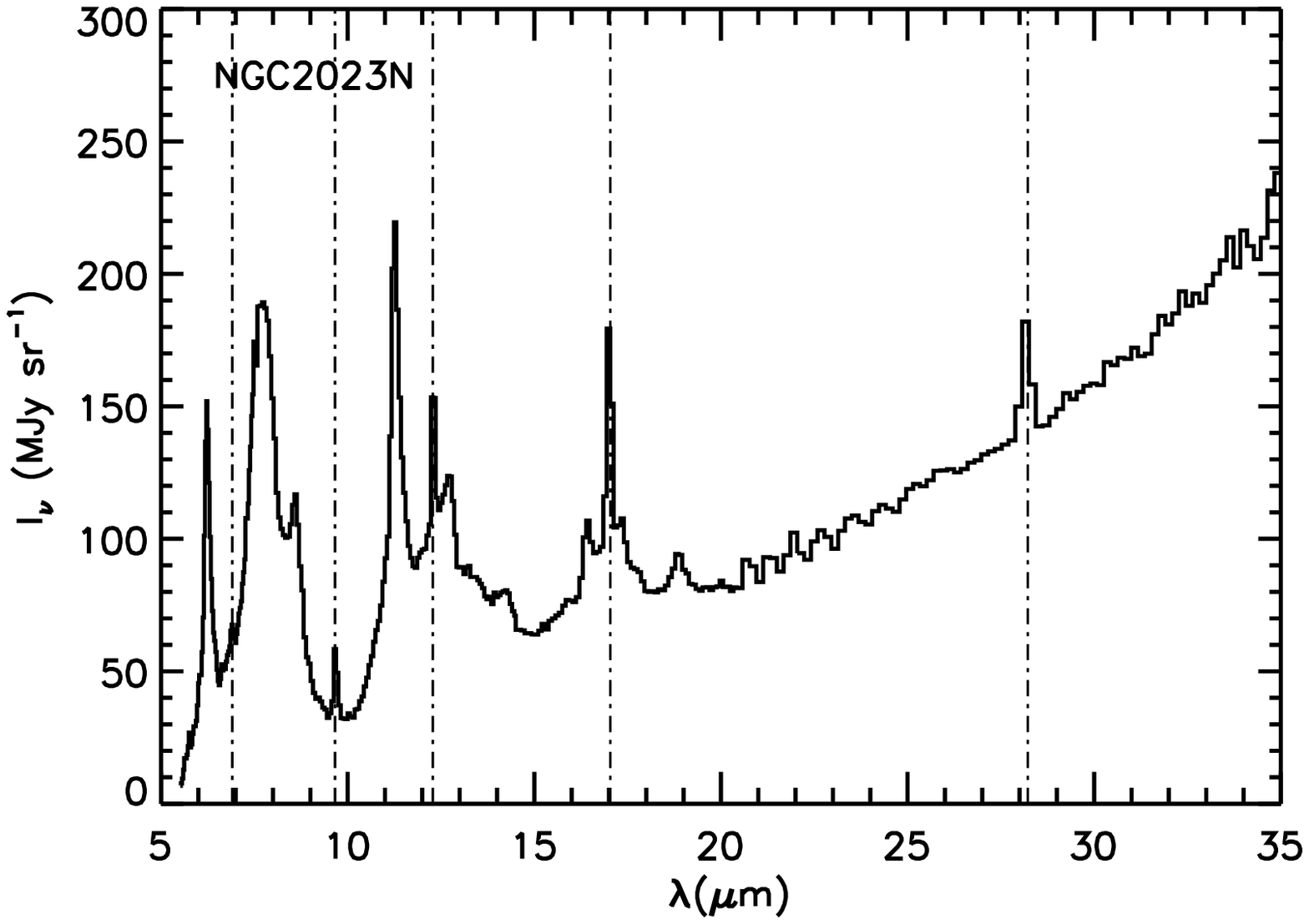}
    \includegraphics[width=0.48\textwidth]{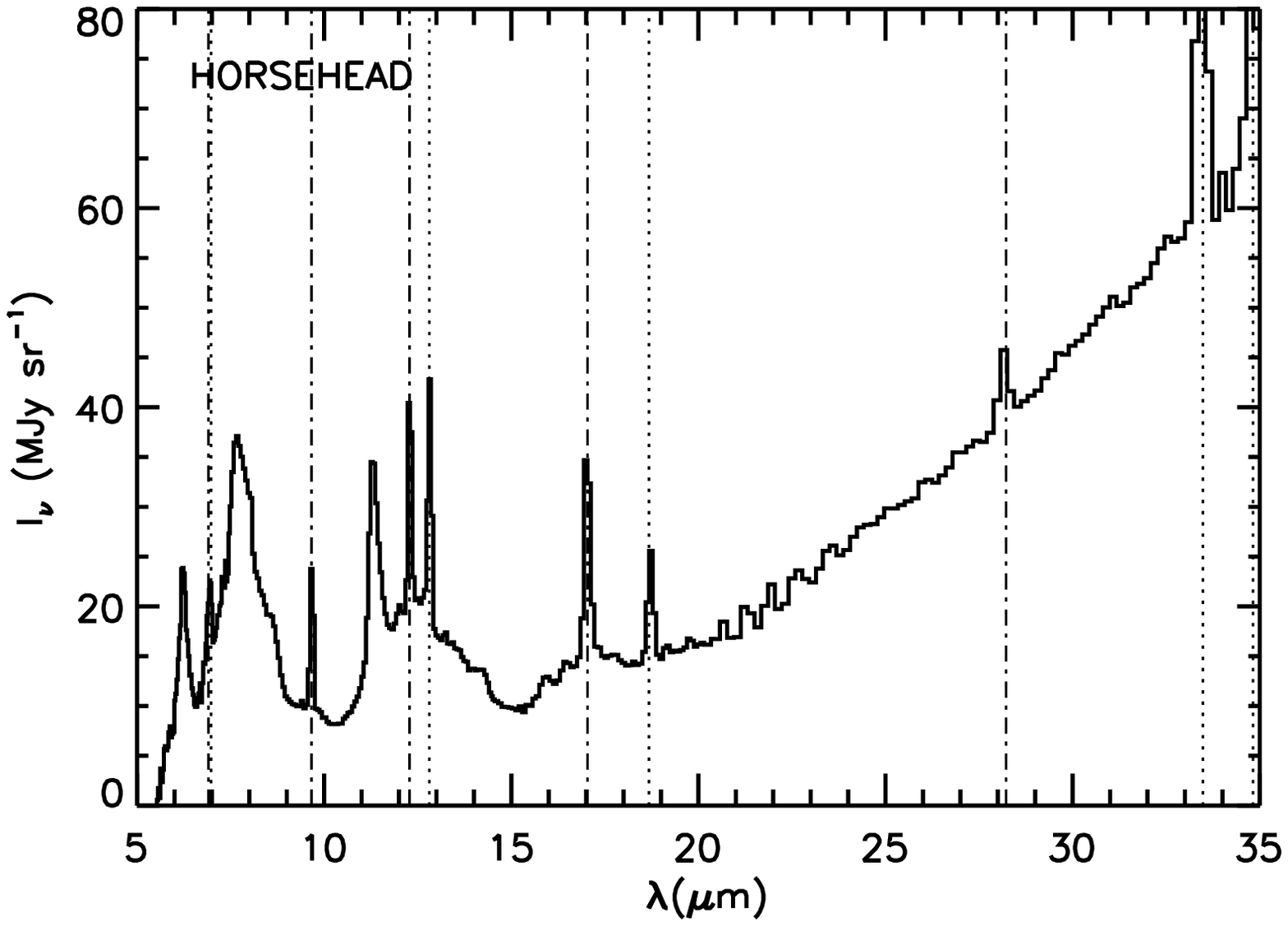}
       \caption{Mean IRS-SL and IRS-LL spectra of NGC2023N and the Horsehead Nebula. The dot-dashed
               lines indicate the wavelengths of the main H$_2$ rotational lines
               0-0\,S(4) to S(0) at 6.9, 9.7, 12.3, 17.0 and 28.2\,$\mu$m which are analyzed elsewhere \citep[][]{habart2008}. 
	       The dotted lines on the 
               Horsehead spectrum indicate the wavelengths of the fine structure lines
              [ArII]6.98\,$\mu$m, [NeII]12.8\,$\mu$m, [SIII]18.7 and 33.4\,$\mu$m, [SiII]
              34.8\,$\mu$m.}
         \label{fig:spectre_moyen}
\end{figure}

\begin{figure*}
   \centering
     \includegraphics[width=0.47\textwidth]{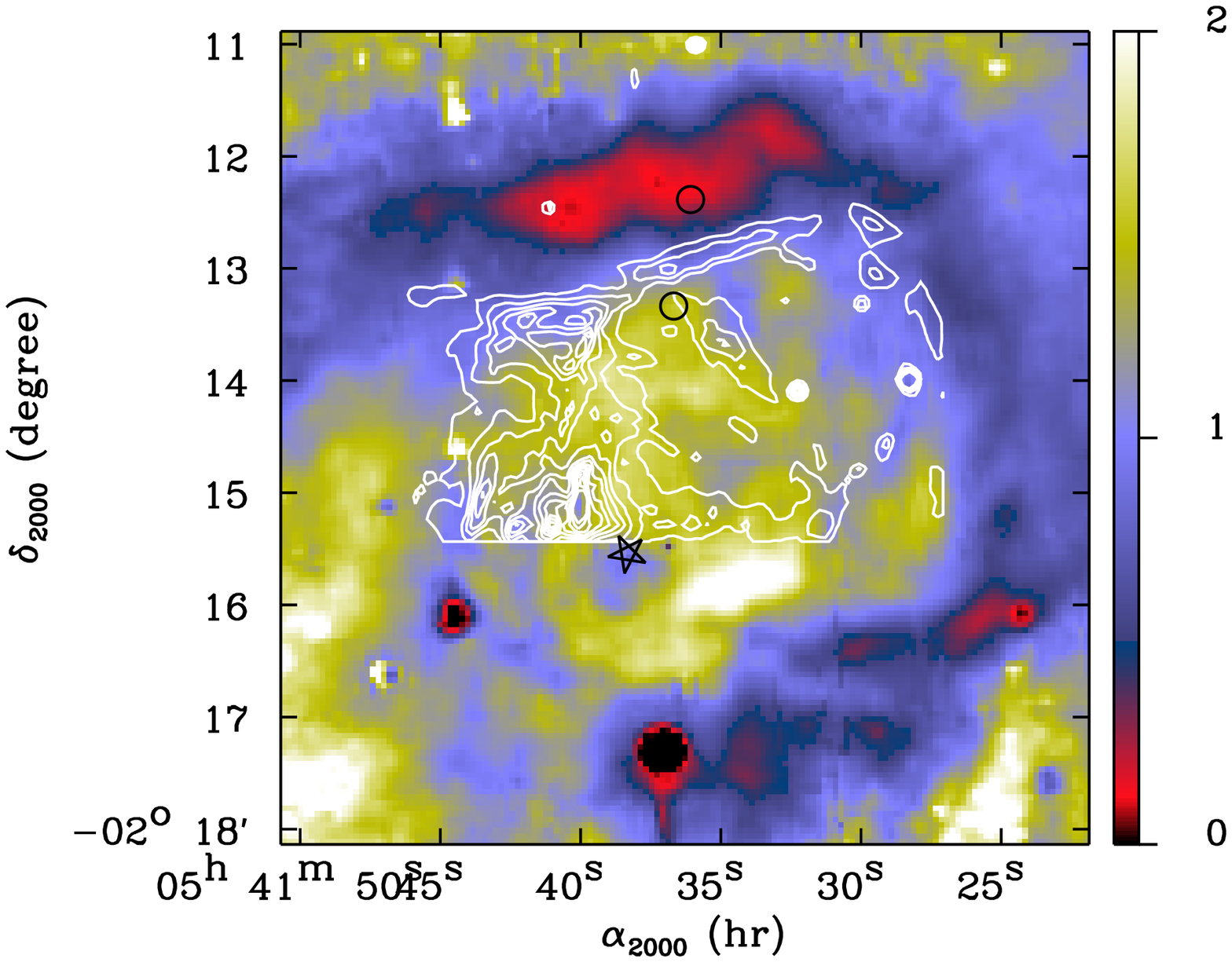}
     \includegraphics[width=0.5\textwidth]{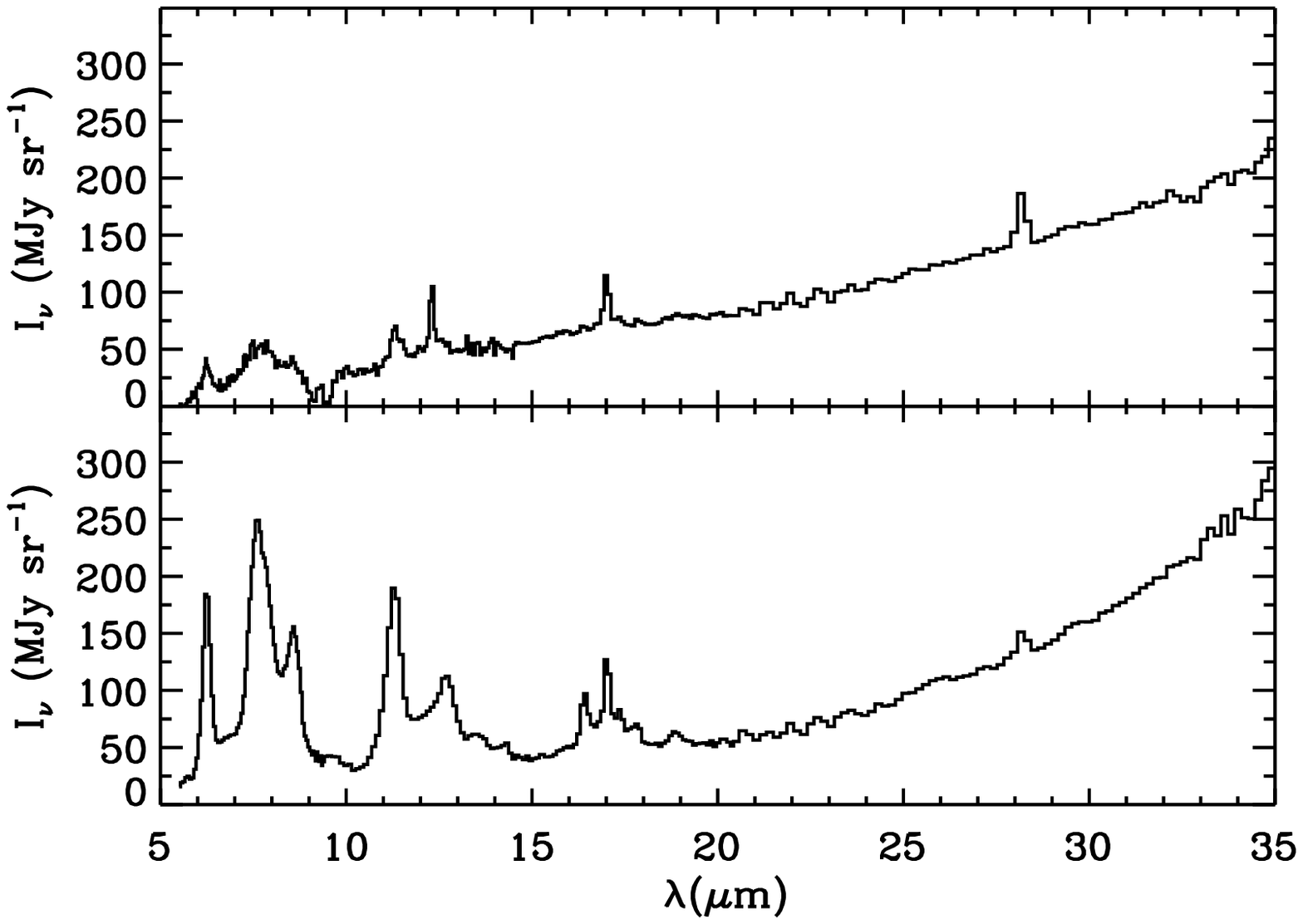}
      \caption{Map of the LW2/LW3
               ratio in NGC2023. The contours are the H$_2$ $\nu$\,=\,1-0\,S(1)
               line at 2.12\,$\mu$m data as seen with the SOFI instrument at
               the ESO New Technology Telescope. 
               The spatial resolution of SOFI data and of LW2
               were degraded (with a Gaussian convolution) to match that of 
               LW3 ($\sim\,$4.5\arcsec).
               The star indicates the position
               of HD37903. The two circles indicate the locations of the spectra
               seen in the right panel.
               The upper spectrum is from the northern position (deep in the cloud) 
               and the lower spectrum
               is from the southern position (close to the star).
               The 5-14.5\,$\mu$m part of the spectra is IRS-SL for the upper spectrum and 
               ISOCAM-CVF for the lower spectrum (and IRS-LL at longer wavelengths in both
                cases).}
         \label{fig:LW2surLW32023}
\end{figure*}

\subsection{ISOCAM data}

Both the Horsehead nebula and NGC2023 were observed with ISOCAM on board the Infrared Space Observatory (ISO) 
in the LW2 (5-8.5\,$\mu$m) and LW3 (12-18\,$\mu$m) broad band photometers
\citep[][]{abergel2002}. The data shown in Fig.\,\ref{fig:OrionB} and Fig.\,\ref{fig:LW2surLW32023} has 
been reduced following \citet{mamd2000}. For the two
filters, the photometric accuracy for the extended emission is
$\sim$\,10\%. The zodiacal contribution subtracted from this
data has also been estimated from the \citet{kelsall98} model.

NGC2023N has also been observed with ISOCAM-CVF
\citep[][]{abergel2002}.  The area observed with ISOCAM-CVF includes
almost entirely that observed with IRS-LL and completely that
observed with IRS-SL. We use this data to obtain the same
spatial coverage for spectral mapping data for the 5-16\,$\mu$m range
as for the 14-35$\mu$m range (IRS-LL). The spectral resolution of
ISOCAM-CVF is $\lambda/\Delta\lambda\,\sim\,$40.
The data reduction and zodiacal light subtraction is described
in \citet{boulanger2005} and the corresponding photometric
accuracy is $\sim$\,20\%.

\section{Spectral variations}\label{sect:result_obs}

Mean mid-IR spectra of both objects are displayed in
Fig.\,\ref{fig:spectre_moyen}.  They are typical of PDRs and present
emission in the AIBs at 6.2, 7.7, 8.6, 11.3, 12.7\,$\mu$m, the mid-IR
dust continuum as well as in H$_2$ rotational lines within the first
vibrational level. The NGC2023N spectrum, which is brighter, also
clearly exhibits the 16.4\,$\mu$m band and the 17.1 and 19\,$\mu$m
bands already reported in NGC7023 by \citet{werner2004b}.  The
spectrum of the Horsehead nebula also presents fine structure lines of
ionised species (dotted lines in Fig.\,\ref{fig:spectre_moyen}) which
arise from the HII region surrounding the Horsehead nebula.  In this
paper, we limit our study to the neutral gas zone \citep[i.e. the PDR,
  see][for study of the HII region]{compiegne2007}.

Fig.\,\ref{fig:LW2surLW32023} shows the map of the band ratio LW2(5-8.5\,$\mu$m)/LW3(12-18\,$\mu$m)
in NGC2023. As already noted by \citet{abergel2002}, this
ratio varies across dense, photodissociated interfaces.
The northern PDR is well delineated by the H$_2$ $\nu$\,=\,1-0\,S(1) 
emission along a nearly east-west ridge at $\delta_{2000}\sim$2\degr13\arcmin\, 
(between the two circle positions in Fig.\,\ref{fig:LW2surLW32023}) which spatially correspond to a 
striking variation of LW2/LW3. 
This ratio indeed decreases by a factor of $\sim$2.6 between the two positions
marked by circles in Fig.\,\ref{fig:LW2surLW32023}
for which spectra are displayed in the same figure.
From their analysis of the ISOCAM-CVF (5-16\,$\mu$m) spectral maps, \citet{abergel2002} concluded that this evolution
of LW2/LW3 is related to a decrease of the AIBs relative to the mid-IR continuum emission while moving away from the star and into 
the molecular cloud. This
conclusion is clearly confirmed by the present IRS spectra which also extend to longer 
wavelengths and better define the mid-IR dust emission. 

The spectra of Fig.\,\ref{fig:LW2surLW32023} also show a variation
of the shape of the 20-35\,$\mu$m continuum between
the two positions. While almost linear deep into the cloud, it shows a non-linear rise closer to the star.
Note that bigger grains which are at thermal equilibrium could be hot enough to emit below
$\lambda\la$\,35\,$\mu$m for positions close to the star while
they are colder deeper in the cloud.

Conversely, the Horsehead observations do not show significant spectral variations. 
This may be due to the sharpness of the photodissociated interface barely resolved by IR data and across 
which the variations occur. 
However, we note that while in NGC2023N the
LW2/LW3 ratio goes from $\sim$1.7 close to the star to $\sim$0.6 deeper into
the cloud, the value of this ratio for the unresolved filament 
of the Horsehead nebula is $\sim$1.1. This value is almost median between the extreme values across NGC2023N. 
In fact, the mean spectra of Fig.\,\ref{fig:spectre_moyen}
have similar LW2/LW3 values ($\sim\,$1) suggesting that we see "average" properties in the Horsehead.

\section{Dust emission model for PDRs}\label{sect:dust_PDR_model}

In order to analyze the IR spectra of PDRs and to trace dust abundance variations
by taking into account excitation effects, 
we developed a dust emission model 
and also treat the radiative transfer problem inside a PDR.
Our dust model is an updated
version of that \citet{desert90} (hereafter DBP).
We combine this model with a 1 dimension radiation transfer model which is fully
consistent with the dust properties of the dust emissivity
model.  Modifications implemented to the DBP model are described in
section\,\ref{sect:dust_model} while the description of the
transfer model is given in section\,\ref{sect:transfer_model}.

\subsection{The dust model}\label{sect:dust_model}

The interpretation of dust observables (extinction and IR emission) in the ISM 
requires at least three dust populations of increasing sizes \citep[e.g. DBP,][]{li2001, Zubko2004}. In DBP, these are:
(i) large (radius $a\sim 0.4$ to 1 nm) polycyclic aromatic hydrocarbons (PAHs) which account for the AIBs 
and for the far-ultraviolet (FUV) non-linear rise in the extinction curve,
(ii) carbonaceous nanoparticles ($a\sim 1$ to 10 nm) called very small grains (VSGs) which carry the mid-IR
continuum emission and the extinction bump at 2175\,\AA\, 
and (iii) big grains (BGs, $a\sim 10$ to 100 nm) which produce the far IR
emission, the $1/\lambda$ rise at visible near-IR 
wavelengths as well as the 10 and 20\,$\mu$m absorption bands in the extinction. In the diffuse interstellar medium, 
small grains (PAHs and VSGs) undergo temperature spikes triggered by the absorption of stellar photons and cool by 
emission in the near and mid-IR range. Conversely, BGs, which have a longer cooling time 
and a shorter timescale between absorption of two photons because of their size, stay at 
constant temperature and emit like grey bodies.

\begin{figure*}
   \centering
     \includegraphics[width=0.49\textwidth]{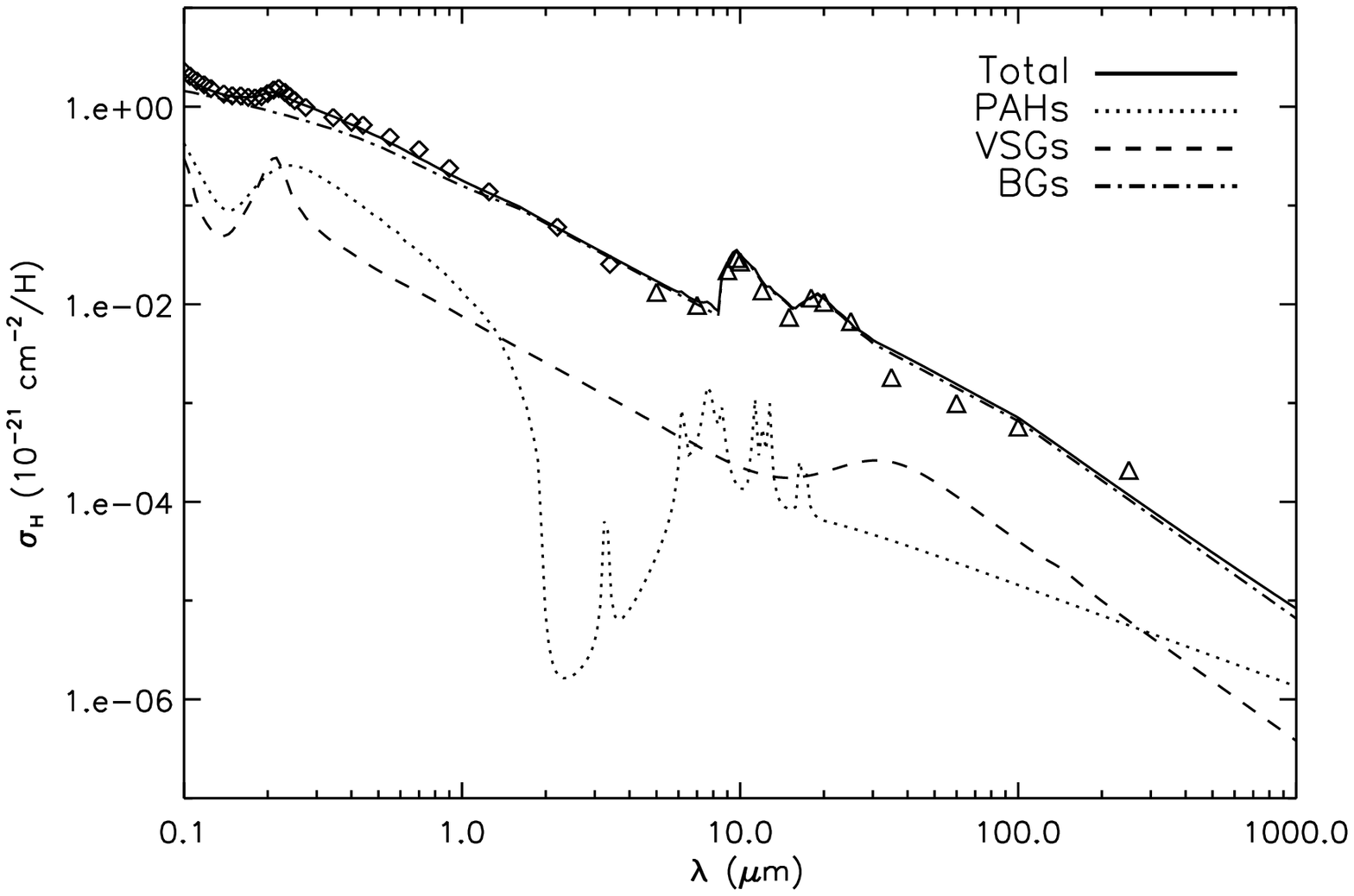}
     \includegraphics[width=0.49\textwidth]{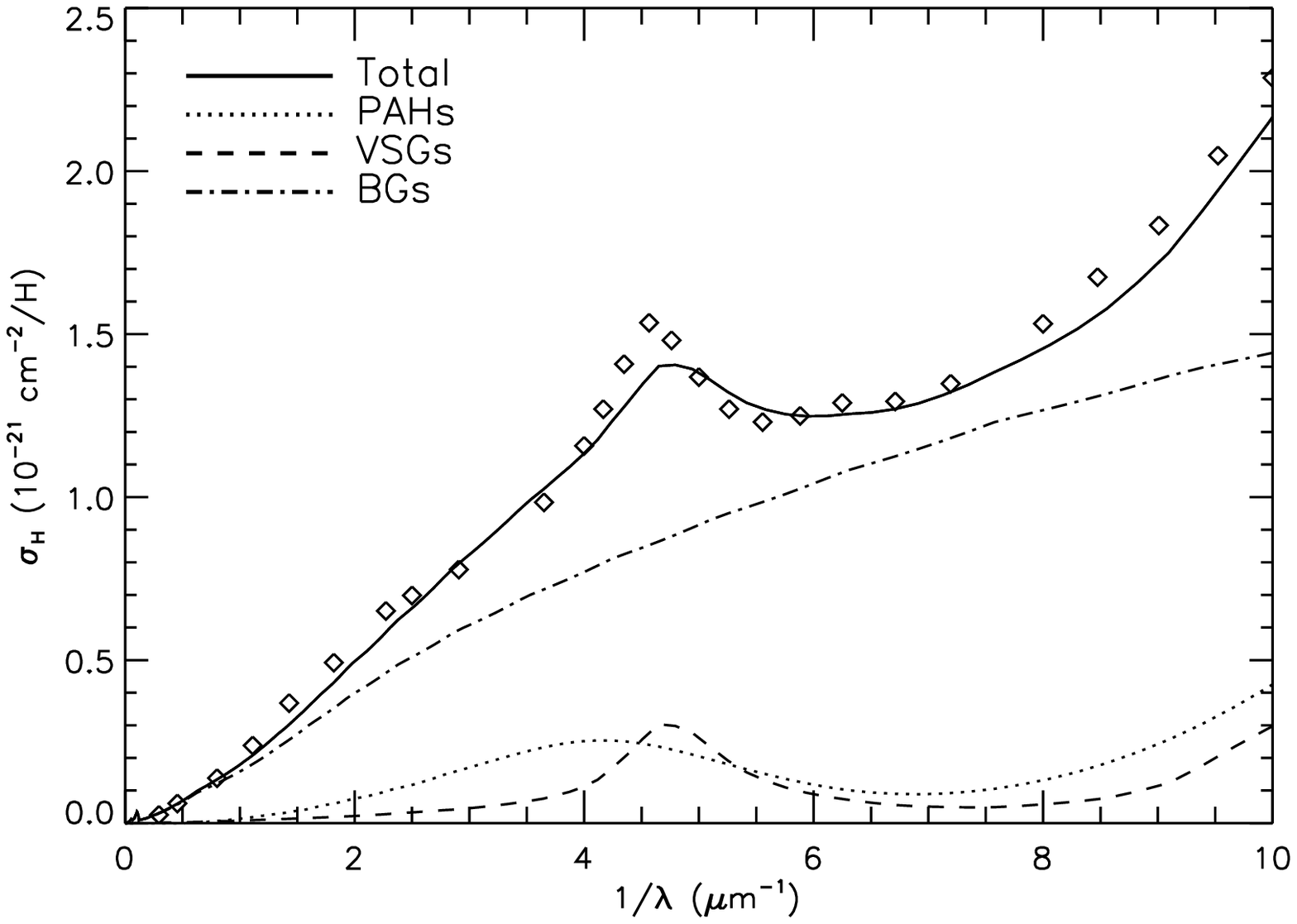} 
     \includegraphics[width=0.49\textwidth]{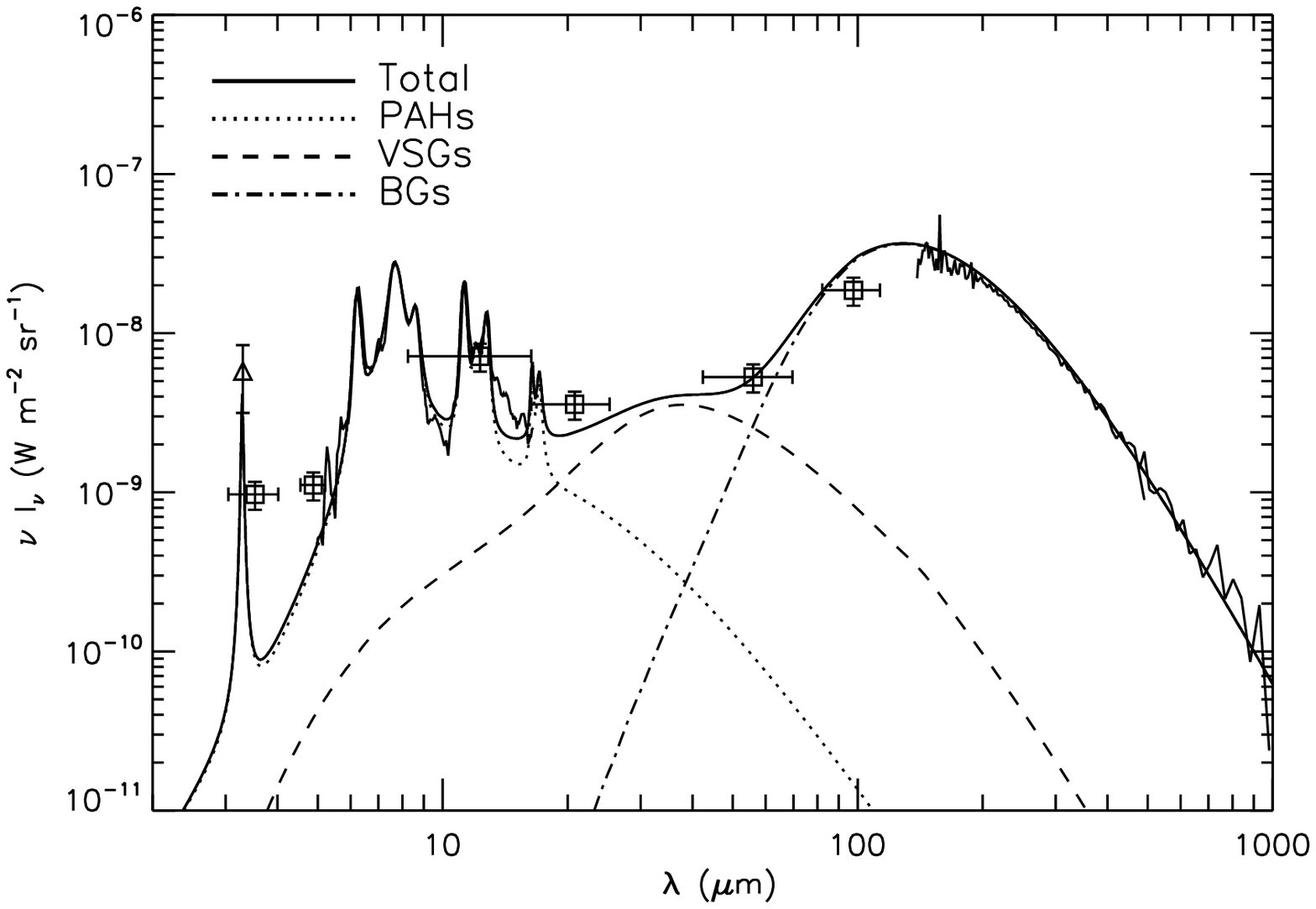}
     \includegraphics[width=0.49\textwidth]{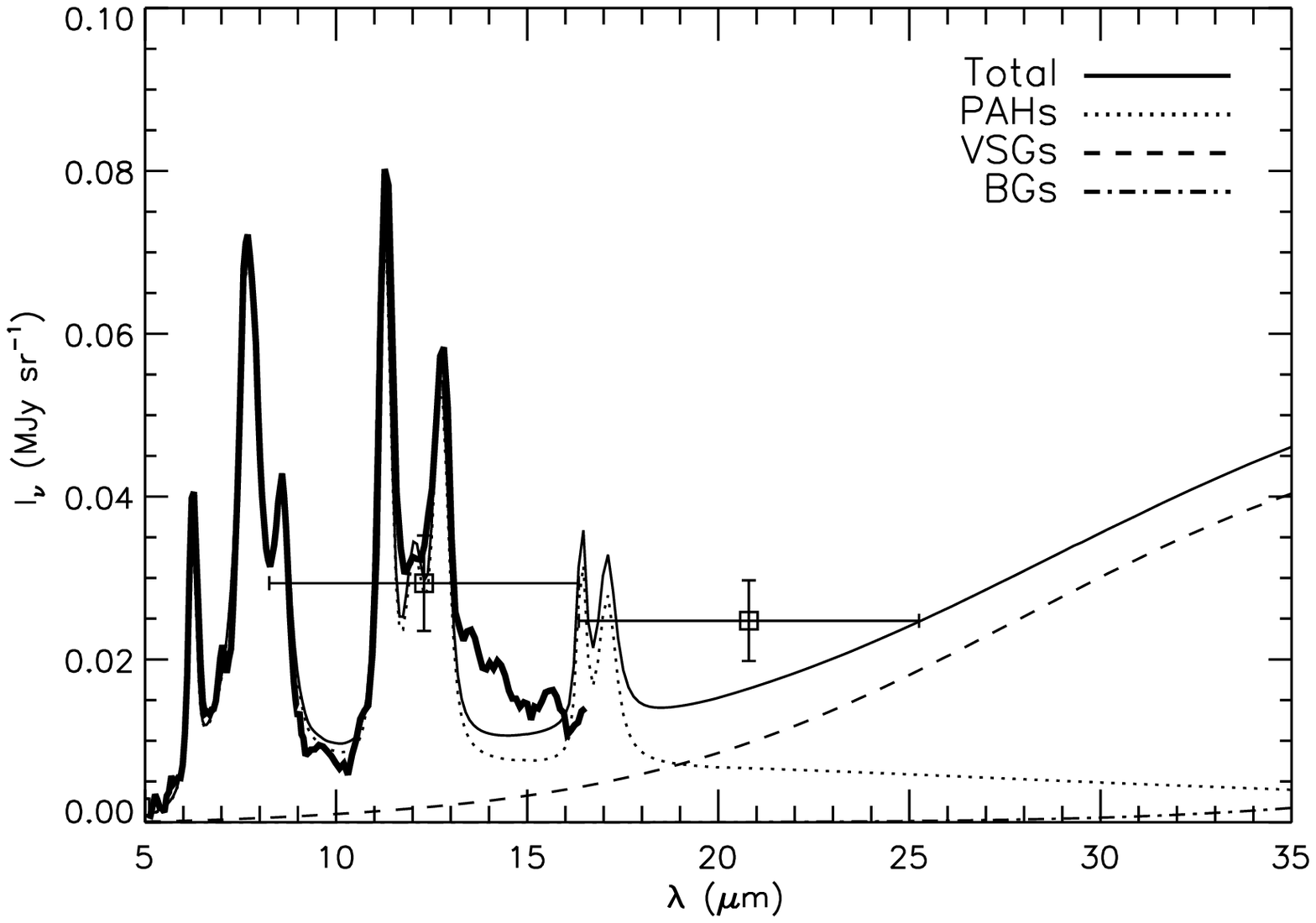}
      \caption{ {\bf Upper panel:} {\it left and right:} Diffuse interstellar medium extinction. The solid line shows our 
        model. Diamonds are the mean
        extinction curve from \citet{savage79} while
        triangles are the mean extinction curve of \citet{mathis90}.
         {\bf Lower panel:} {\it left} Emission spectrum for the diffuse interstellar
        medium for $N_H\,=\,10^{20}\,cm^{-2}$.  
        The solid line is our dust model. Data points are 
        (i) the 3.3\,$\mu$m emission (triangle) of cirrus associated with
        the molecular ring \citep[][]{giard94}, (ii) the ISOCAM-CVF spectrum
        between 5 and 16\,$\mu$m of the diffuse Galactic emission at
        b$\sim$1\degr \citep[][]{flagey2006}, (iii) the DIRBE
        measurements (squares) for $|b|>25$\degr \citep{arendt98} and (iv) the FIRAS submm spectrum 
        of \citet{boulanger96}. 
        {\it Right:} Diffuse ISM model emission spectrum (thin solid line) over the
        wavelength range of IRS. The ISOCAM-CVF spectrum is superimposed (thick solid line). The
        model spectrum has been smoothed to the resolving power (R$\sim$40) of ISOCAM for $\lambda<15\,\mu$m.}
         \label{fig:diffuse_dust_spectra}
\end{figure*}

The dust properties that we describe below allow the reproduction of the extinction curve and emission
spectrum of the diffuse high galactic latitude ISM (so called "Cirrus") 
as seen in
Fig.\,\ref{fig:diffuse_dust_spectra}. The exciting radiation 
is the Interstellar Standard Radiation Field \citep[ISRF,][]{mathis83}.
We call this dust population {\it Cirrus dust} as a
reference to compare to dust emission in other interstellar environments.
  
The absorption cross-section of PAHs has been updated. We take the visible-UV cross-section from 
\citet{verstraete92} and apply their size dependent cut-off for electronic transitions in the near-IR.
The resulting cross-section provides a good match to the available 
laboratory measurements on species containing 20 to 30 C atoms \citep{joblin92}.
The cross-sections of the AIBs (at 3.3, 6.2, 7.7, 8.6, 11.3, 12.7 and 16.4\,$\mu$m) are based on \citet{verstraete2001} 
and \citet{pech2002} and assumes singly ionized PAHs. ISO-SWS spectra also indicates the presence of a broad band at 
6.9\,$\mu$m and a band at 12\,$\mu$m \citep{verstraete2001, peeters2002a}. In addition, Spitzer data has shown the presence 
of another band at 17\,$\mu$m which behaves like the rest of the AIBs \citep{peeters2004,smith2007}.
We therefore include these bands in our AIB list which is summarized in Table\,\ref{tab:AIBs}. The positions and widths of these bands were derived from ISO and Spitzer spectra.
Peak values of cross sections of all AIBs are adjusted in order to match the Cirrus observed spectrum.
%
%-----------------------------------------------------------------------------------------------
\begin{table}[t]
  \centering
  \begin{tabular}{c c c c}
   \hline           
   \hline           
        $\lambda_i$ & $\nu_i$      & $\Delta\nu_i$ &  $\sigma_i$ \\
         ($\mu$m)   & ($cm^{-1}$)     &   ($cm^{-1}$)       &    (10$^{-20} cm^{2}$) \\
   \hline
   3.3        &    3040    &   39     &   2.4 $N_H$      \\

   6.2        &    1609    &   44     &   3.1 $N_C$     \\

   6.9        &   1450     &   300    &   0.4 $N_C$     \\

   7.7        &    1300    &   113   &   3.8 $N_C$     \\ 

   8.6        &    1162    &   47     &   4.9 $N_H$     \\ 
 
   11.3       &    890    &   18     &   11.0 $N_H$     \\ 
 
   12.0       &    830    &   30     &   3.2 $N_H$     \\ 
 
   12.7       &    785    &   16     &   7.5 $N_H$     \\ 
 
   16.4       &    609    &   6     &   1.0 $N_C$    \\ 

   17.1       &    585     &   17    &  0.6  $N_C$  \\   
   \hline
  \end{tabular}
  \caption{Parameters of the AIBs with $N_C$ and $N_H$
           the number of carbon and hydrogen atoms in the molecule respectively.
	   CC(CH) bands strength scale with $N_C$ ($N_H$).
           }
  \label{tab:AIBs} 
\end{table}
%-----------------------------------------------------------------------------------------------
%-----------------------------------------------------------------------------------------------
\begin{table}[t]
  \centering
  \begin{tabular}{c c c c}
   \hline           
   \hline           
               & $\alpha $ & $a\rm{_{min}}$ & $a\rm{_{max}}$  \\
               &           &    $nm$     &  $nm$        \\
   \hline
   PAH         &    3.5    &   0.4     &   1.2      \\
   \hline
   VSG         &    3.5    &   1.0     &   4.0      \\
   \hline
   BG          &    3.5    &   4.0     &   110      \\ 
   \hline
  \end{tabular}
  \caption{Parameters of the size distribution n($a$) for the three components of our dust model.
           We assume that n$(a)\sim a^{-\alpha}$ with $a$ between $a_{\rm min}$ and $a_{\rm max}$.
          }
  \label{tab:size_DUSTEM} 
\end{table}
%-----------------------------------------------------------------------------------------------
%
We assume planar and fully hydrogenated PAHs with hexagonal symmetry ($D_{6h}$). 
The molecular radius $a$ is related to the number of carbon atoms per molecule $N_C$
as $a(\AA)\,=\,0.9\sqrt{N_C}$ and the number of hydrogen atoms is given by $N_H\,=\,\sqrt{6\,N_C}$.
Finally we take the far-IR cross section from \citet{schutte93}.

VSGs are assumed to be graphitic and both the 
optical properties and heat capacity have been updated \citep[e.g.][]{Draine2001}.
Their optical properties have been calculated with Mie
theory\footnote{Code retrieved from
  http://atol.ucsd.edu/scatlib/index.htm} applied to small graphitic
spheres in the 2/3\,-\,1/3\, approximation
\citep[e.g.][]{draine84}. The dielectric constant was taken from \citet{laor93}.

As in DBP, BGs are taken to be silicates with carbonaceous mantles or inclusions in order to provide the required
absorption efficiency in the NIR compared to bare silicates.
The near-IR absorption cross section of BGs now has a $\lambda^{-1.5}$ dependence 
which allows a better match of the observed extinction curve.

We assumed the size distributions of all the dust populations of our model 
to be a power-law type, $n(a)\propto a^{-3.5}$ with minimum and maximum radii
$a_{\rm min}$ and $a_{\rm max}$ listed in Table\,\ref{tab:size_DUSTEM}. 
This simple form of size distribution originally introduced by \citet{MRN1977} 
remains a good paradigm \citep[][]{ kim94a, Weingartner2001, Clayton2003}.
Note that abundances of the three populations remain unchanged from
those of the DBP model.

One of the consequences of the update is that UV-visible cross sections 
(see Fig.\,\ref{fig:diffuse_dust_spectra}) of the PAHs
and VSGs are more similar than in the DBP model. 
In fact, these two species are assumed to contain mostly sp$^2$
hybridization carbons which are responsible for the 2175\,$\AA$ bump
($\rm{\pi\,\rightarrow\,\pi^{\ast}}$ electronic transition) and for
the non-linear rise in the far UV (bump at 800\,$\AA$ due to a
$\rm{\sigma\,\rightarrow\,\sigma^{\ast}}$ electronic transition).
This similarity has consequences for the possible self shielding of the
two species which will cause an evolution of the spectral shape
depending the shape of the exciting radiation field spectrum.

\subsection{Radiative transfer}\label{sect:transfer_model}

\begin{figure}
   \centering
     \includegraphics[width=0.47\textwidth]{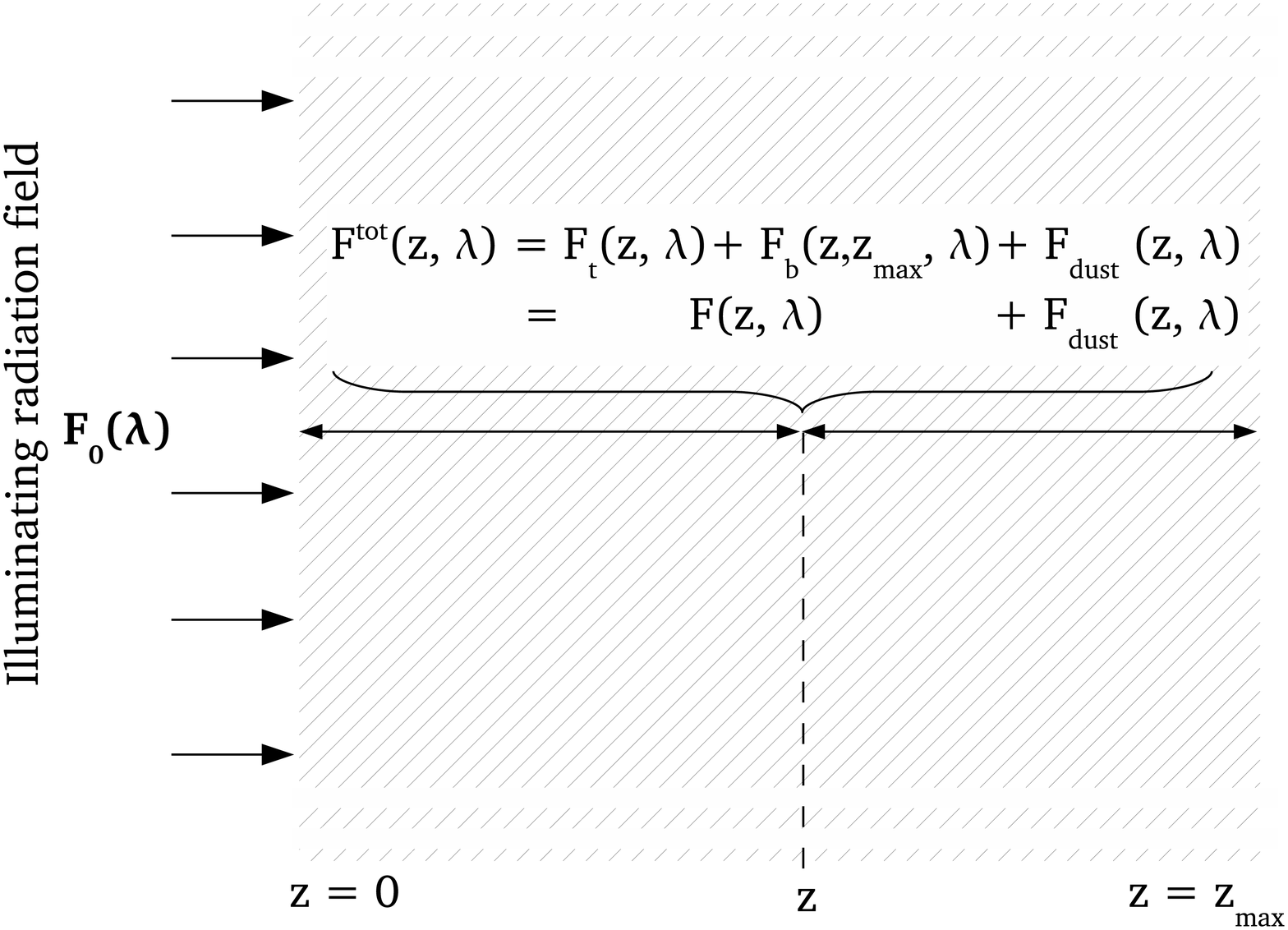}
      \caption{Schematic view of the plane-parallel geometry used. 
               $F^{\rm{tot}}(z, \lambda)$ is the total exciting flux which is considered in
               deriving the dust emissivity. $F_{\rm{dust}}(z, \lambda)$
               is the flux related to the dust emission from the cloud itself. 
               $F_{t}(z, \lambda)$ and $F_{b}(z,z_{\rm{max}}, \lambda)$ are the flux
               related to the illuminating radiation field transmitted
               from the illuminated edge and backscattered by the deeper part of the cloud, respectively.}
         \label{fig:transfer_pp}
\end{figure}

We describe here how the total exciting flux $F^{\rm{tot}}(z,
\lambda)$ is estimated at each depth $z$ of the cloud.  This total
exciting flux results from the contribution of the illuminating
radiation field transfered into the cloud $F(z, \lambda)$ and of the
dust emission $F_{\rm{dust}}(z, \lambda)$, as illustrated by
Fig\,\ref{fig:transfer_pp}.

We first transfer the incident radiation field into the cloud which is
represented by a semi-finite plane-parallel slab of proton density
$n_H(z)$ at each depth $z$ ($z=0$ at the free interface).  We use the
dust properties described above. Note that the PAHs are assumed to have a
zero albedo and that the albedo of VSGs is low due to their small size. The
scattering cross section is thus dominated by the contribution of BGs.

The incident radiation field $F_0$($\lambda$) is the sum of the mean
interstellar radiation field (ISRF) of \citet{mathis83} and of a
blackbody whose flux is diluted by ($R_{\star}/d)^2$ with $d$ the
current distance to the star and $R_{\star}$ the stellar radius.  Note
that the flux of the ISRF is obtained by integrating the intensity
over 2$\pi$ steradians since the cloud is illuminated on one
side. Moreover, we limit the radiation field to photon energies of
less than $13.6\,eV$ since we consider neutral gas (PDR).

The plane-parallel approximation allows us to simplify the treatment of 
the scattering. We assume that each incident photon is either scattered forward
deeper into the cloud (increasing $z$) or scattered back towards the cloud interface (decreasing $z$).
Assuming that a photon can only be backscattered once, the flux of the 
radiation field at a given depth $z$ into the cloud can be written as
\begin{equation}
F(z,\lambda)\,=\,F_{t}(z,\lambda)\,+\,F_{b}(z,z_{\rm{max}}, \lambda)
 \label{eq:illuminating_flux}
\end{equation}
where $F_{t}(z,\lambda)$ is the transmitted flux
from the illuminated surface of the cloud and $F_{b}(z,z_{\rm{max}}, \lambda)$
the flux backscattered by the part of the cloud deeper than $z$. 
We impose a finite depth $z\rm{_{max}}$ for the cloud in order
to estimate the backscattered flux.

With our assumptions, we only need absorption cross sections ($\sigma_{a}(\lambda)$) and 
backscattering cross sections ($\sigma_{b}(\lambda)$)
to perform the transfer since
photons that are scattered forward are simply treated as transmitted.
Our dust model only computes the integrated scattering
($\sigma_{s}(\lambda)$)
cross section and not the differential one ($d\sigma_{s}(\lambda)\,/\,d\Omega$). We thus use
the \citet{henyey41} phase function ($\Phi_{\rm{H-G}}(\theta)$) to calculate the 
backscattering cross section as
\begin{equation}
\sigma_{b}(\lambda)\,=\,2\pi\,\sigma_{s}(\lambda)\,\int_{\pi/2}^{\pi}\,\Phi_{\rm{H-G}}(\theta)\,sin\theta\,d\theta.
\end{equation}
Note that when taking
$g\,=\,<cos(\theta)>\,=\,0.6$ which is the common value used, 
in accordance with the diffuse interstellar medium observations \citep[e.g.][]{witt97, schiminovich2001},
$\sim 12\,\%$ of scattered flux is backscattered.

The transmitted flux is given by
\begin{equation}
F_{t}(z,\lambda)\,=\,F_0(\lambda)\,e^{-\tau_{\rm{ext}}(0\rightarrow z, \lambda)}
\end{equation}
with the opacity $\tau_{\rm{ext}}(0\rightarrow z, \lambda)$ defined as
\begin{equation}
\tau_{\rm{ext}}(z_1\rightarrow z_2, \lambda)\,=\,\int_{z1}^{z2}\,n_H(z')\,(\sigma_{a}(z',\lambda)\,+\,\sigma_{b}(z',\lambda))\,dz'
 \label{eq:tau_ext}
\end{equation}
The backscattered flux $F_{b}(z,z_{\rm{max}}, \lambda)$ is given by
\begin{equation}
F_{b}(z,z_{\rm{max}},\lambda)\,=\,
       \int_z^{z_{\rm{max}}} \,dF_{b}(z',\lambda)\,
       e^{-\tau_{\rm{ext}}(z'\rightarrow\,z,\,\lambda)}
\end{equation}
with
\begin{equation}
dF_{b}(z',\lambda)\,\approx\,F_{t}(z',\,\lambda)\,{d\tau_{b}(z',\,\lambda)}
\end{equation}
where $d\tau_{b}(z',\,\lambda)$ is the backscattering optical depth of the layer $dz'$ at a depth $z'$.

From the exciting flux $F(z, \lambda)$ due the illuminating radiation field, 
we derive the dust emissivity $\epsilon'(z,\lambda)$ which will be used 
to calculate the exciting flux $F_{\rm{dust}}(z, \lambda)$ due to the dust emission.
In the plane-parallel approximation, we can estimate this contribution as
\begin{equation}
F_{\rm{dust}}(z, \lambda)\,=\,\frac{1}{2}\,\int_0^{z_{\rm{max}}}\,n_H(z')\,\epsilon'(z', \lambda)\,e^{-\tau_{\rm{ext}}(z'\rightarrow z,\,\lambda)}\,dz'
\end{equation}
with $\epsilon'(z', \lambda)$ the dust emissivity resulting from the
illuminating radiation field $F(z, \lambda)$, $n_H(z')$ the density
profile and $\tau_{\rm{ext}}(z'\rightarrow z,\,\lambda)$ the
extinction opacity between the depth $z'$ and $z$ as defined by
eq.\,\ref{eq:tau_ext}. The 1/2 factor is related to the fact that the
dust emission is assumed to be isotropic and that consequently, half
the power emitted at a depth $z'$ does not reach the depth $z$, since
the dust scattering cross section is very low at the relevant
wavelengths.  Note that for the studied depth ($A_v\,\la\,20$), an
iterative computation of the dust emission contribution to the dust
heating is not required since the mid and near IR dust emission, which
contributes significantly to the dust heating is mostly emitted by the
dust at low depth, whose heating is completely dominated by the
UV-visible photons of the illuminating radiation field.

We can finally derive the dust emissivity $\epsilon(z,\lambda)$ as a function of the depth
in the cloud, as excited by $F^{\rm{tot}}(z, \lambda)\,=\,F(z, \lambda)\,+\,F_{\rm{dust}}(z, \lambda)$.

\section{Modelling the Horsehead nebula}\label{sect:model_HH}

\begin{figure*}
   \centering
     \includegraphics[width=0.47\textwidth]{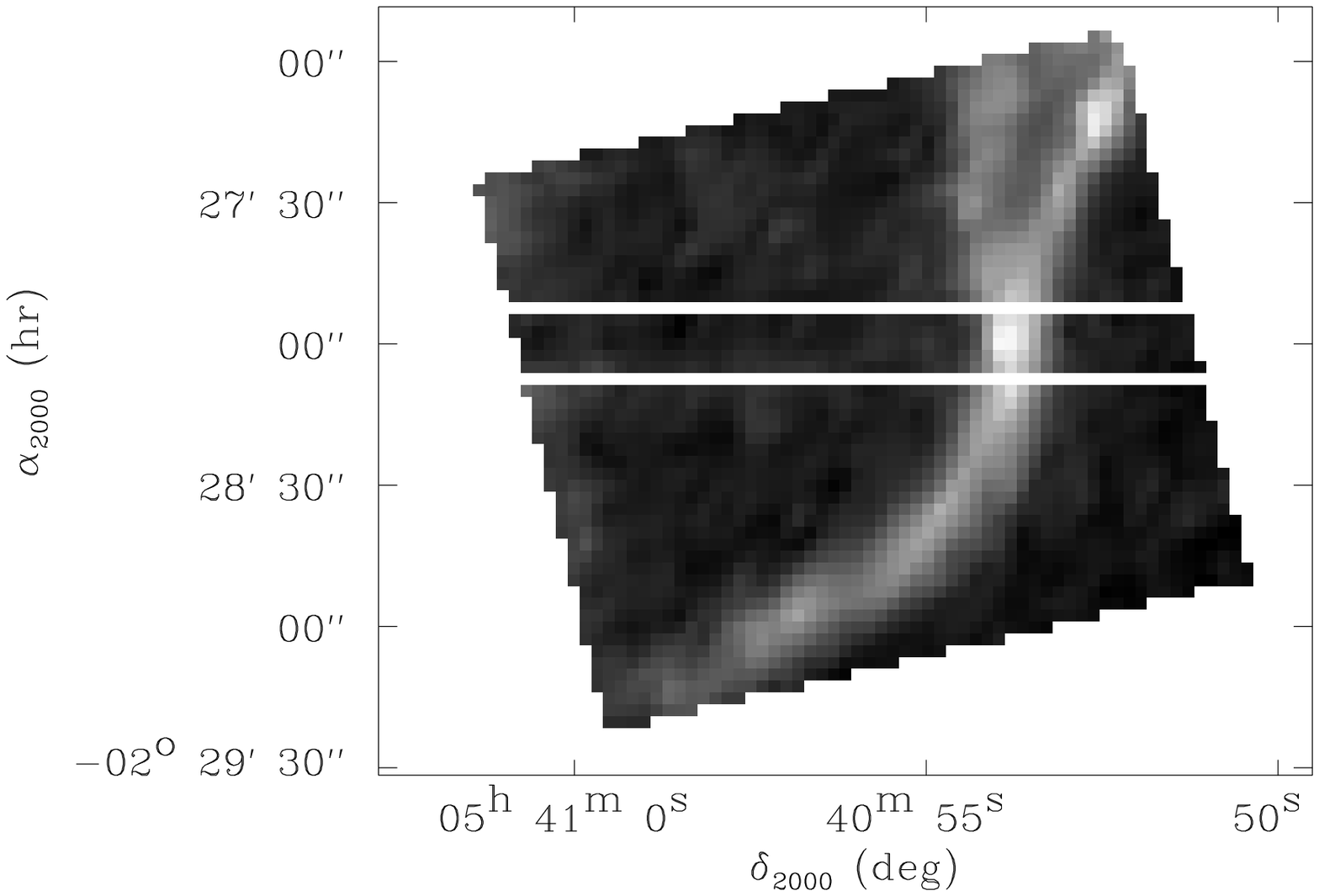}
     \includegraphics[width=0.47\textwidth]{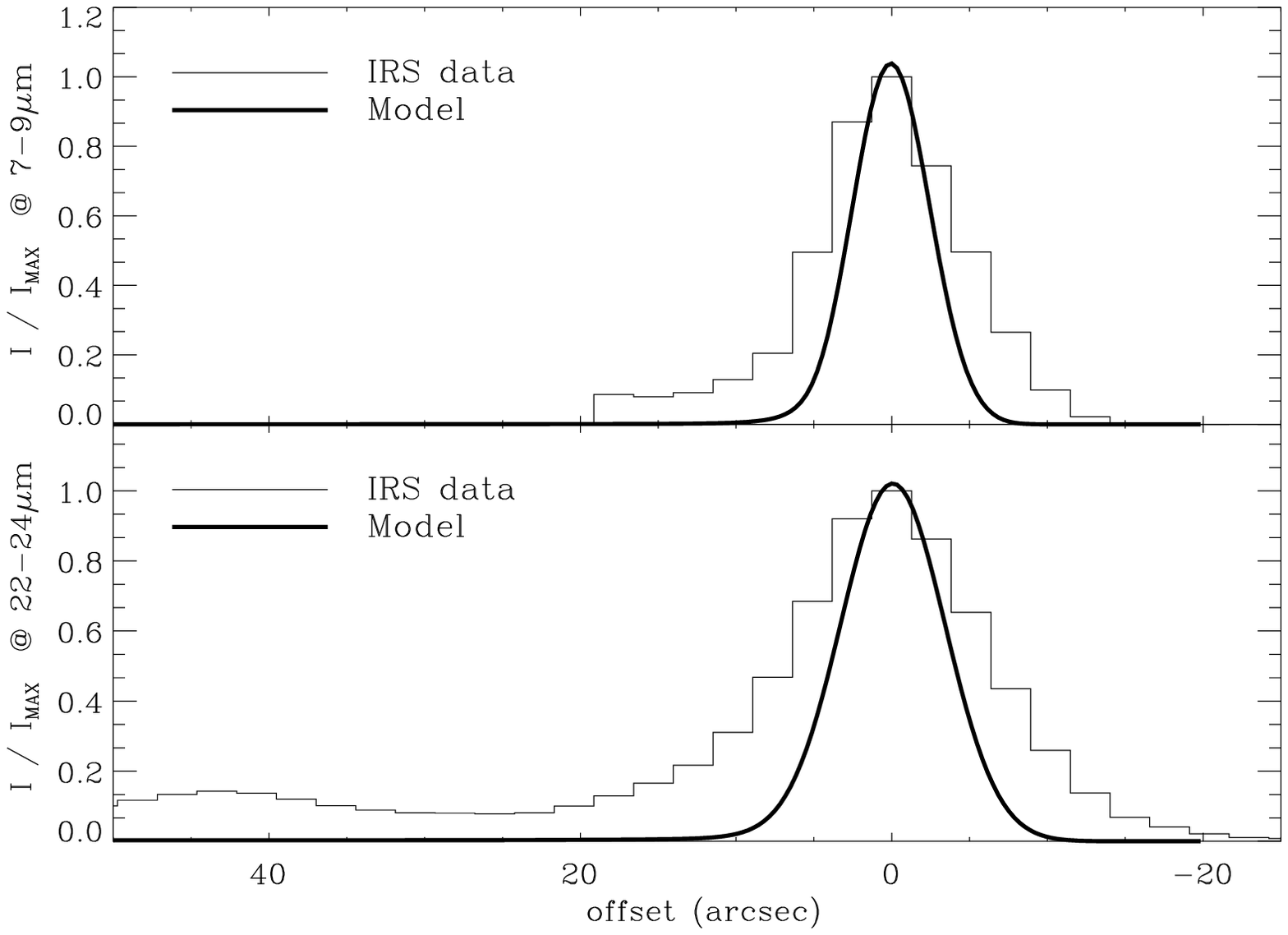}
      \caption{{\bf Left panel:} Image of the Horsehead Nebula has seen by IRS-LL. The two cuts show
               the limit between which the mean observed profiles are computed. 
               {\bf Right panel:} Comparison of the observed and modelled profiles at 7-9\,$\mu$m (AIBs) and
               22-24\,$\mu$m (mid-IR continuum) normalized to the peak values. The offset origin of the cuts
               correspond to the 7-9\,$\mu$m emission peak. $\sigma$Orionis is located toward negative offsets.}
         \label{fig:model_profil}
\end{figure*}

Due to its almost edge-on geometry, the Horsehead nebula can be
modelled using plane-parallel geometry.

The radiation field is characterized by
 $R_\star\,=\,8.5\,R_\odot$
and $T_{eff}\,=\,34620\,K$ corresponding to an O9.5V star \citep[][]{Schaerer97}
located at a distance of 3.5\,pc from the cloud.
We use the density profile deduced by \citet{habart2005}
from a comparison of model gas emission profiles to observations of H$_2$, CO and dust emission:
\begin{equation}
n_H(z) = \left\{
    \begin{array}{lll}
         n_H^0\,\times\,(z/z_0)^\beta & : & z\,\leq\,z_0 \\
         n_H^0                       & : & z\,>\,z_0
    \end{array}
\right.
\end{equation}
with $n_H^0\,=\,2\,10^5\,cm^{-3}$, $z_0\,=\,0.02\,pc$ 
(10\arcsec) and $\beta$\,=4.
We integrate the output emissivity $\epsilon(\lambda, z)$ 
along the line of sight to obtain intensities in units of $\rm{MJy\,sr^{-1}}$, those of our data.
We further impose a length for the PDR along the line of sight, namely, $l\rm{_{PDR}}$.
We neglect the absorption at these wavelengths since the visual extinction
range between $A_V\sim\,2\,-\,30$ as function of $z$ in the line of sight along the Horsehead.
As in \citet{habart2005}, we perform a 6\degr\, rotation of the PDR
to the line of sight.

\subsection{Emission profiles}

Fig.\,\ref{fig:model_profil} shows observed and modelled profiles
at 7-9\,$\mu$m and 22-24\,$\mu$m normalized to the peak value. 
These spectral ranges are unambiguously related to the AIBs (7-9\,$\mu$m) and the mid-IR continuum
(22-24\,$\mu$m) which was not the case for the ISOCAM-LW3 (12-18\,$\mu$m) 
broad band which contains both mid-IR continuum and certain AIBs (the 12.7 16.4, 17.1\,$\mu$m).
Modelled profiles were convolved by a Gaussian whose width corresponds
to the IRS spatial resolution at the corresponding wavelengths (1.9\arcsec at 7-9\,$\mu$m and 5.5\arcsec at 22-24\,$\mu$m).
The observed profile is defined as the mean over the strip
shown in Fig.\,\ref{fig:model_profil}.
Both the observed and modelled angular offsets are defined 
with respect to the peak emission position at 7-9\,$\mu$m.
We can then see that the mid-IR continuum (22-24\,$\mu$m)
and AIB (7-9\,$\mu$m) emission peak 
at the same depth in the cloud and are well reproduced by the 
model. 

Although we take into account the profile broadening 
caused by the point spread function and by the tilt
of the illuminated ridge, we see
that the modelled profiles are narrower than the observed one.
This must be related to differences between our plane-parallel representation
and the real geometry of the Horsehead nebula (e.g. multiple filaments
superimposed).
The absolute intensity of the mid-IR continuum (22-24\,$\mu$m) at the peak position can be reproduced
if $l\rm{_{PDR}}$\,=\,0.08\,pc. 
This value is only 20\% lower than that found by \citet{habart2005}.
However, the modelled AIB intensity is then 
2.4 times higher than the observed one (i.e would have required $l\rm{_{PDR}}$\,=\,0.033\,pc
for the absolute intensity to be reproduced). We can then conclude 
that taking the radiative transfer effect into account, "Cirrus dust" properties
cannot explain the observed AIBs/mid-IR continuum ratio.

\begin{figure*}
   \centering
     \includegraphics[width=0.47\textwidth]{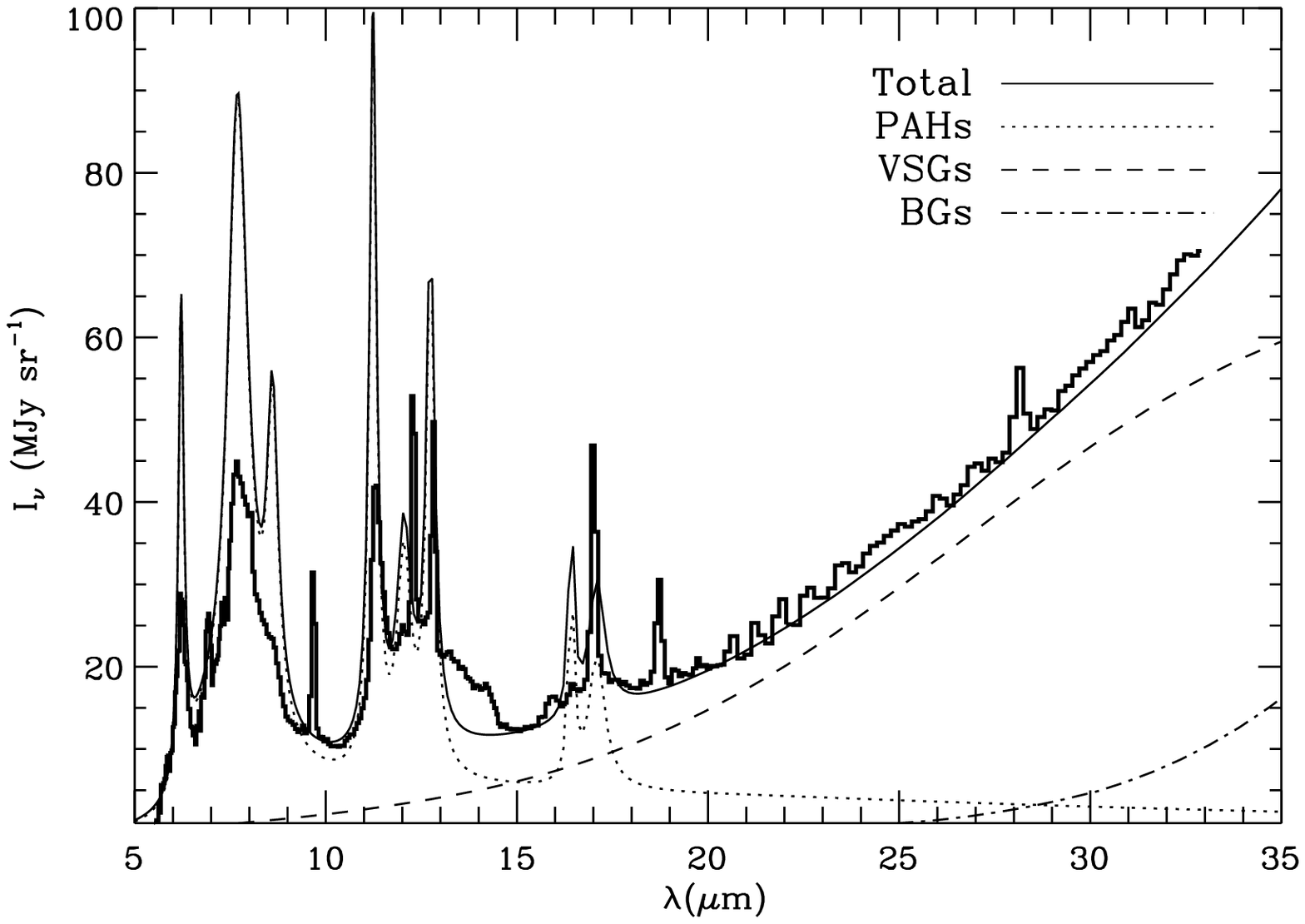}
     \includegraphics[width=0.47\textwidth]{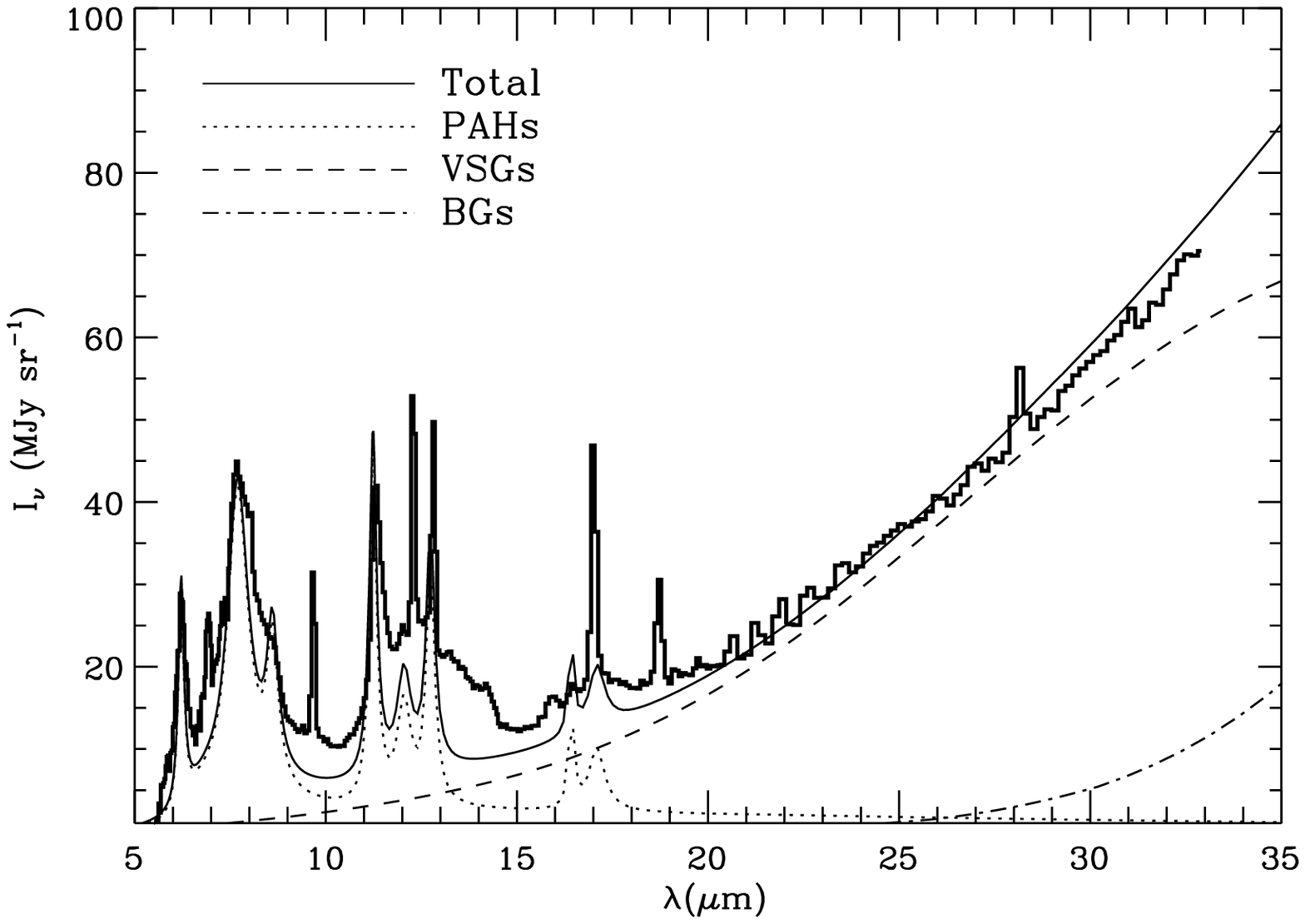} 
      \caption{ Comparison between observed (thick line) and modelled (thin line) spectra at the peak position of the emission
                of the Horsehead Nebula
                (see \S\,\ref{sect:comparison_spectra_HH} for the exact definition). 
                In the {\bf left panel}, we show Cirrus dust emssion.
                In the {\bf right panel}, the PAH abundance has been reduced by 2.4 with respect to the Cirrus abundance.
                Note that IRS spectra have been cut at $\lambda\,\sim\,33\,\mu$m to avoid the fine-structure lines [SIII]\,33.4\,$\mu$m 
                and [SiII]\,34.8\,$\mu$m.}
         \label{fig:spectre_model_HH}
\end{figure*}

\subsection{Spectral comparison}\label{sect:comparison_spectra_HH}

The AIBs excess with respect to the mid-IR continuum can also be seen
in Fig.\,\ref{fig:spectre_model_HH}. 
In both model and observations, spectra of this figure are defined as the average over 
positions whose intensities at 7-9\,$\mu$m are at least 2/3 of the peak values.
For the observed spectra, we must ensure that we
are looking at the same position within the cloud (i.e. same depth) since we are looking at 
wavelengths ranging from 5 to 33\,$\mu$m whose spatial resolution are not the same. 
We therefore bring the resolution at all wavelengths to 7.9\arcsec, i.e., that of the 33\,$\mu$m
by convolving the spectra with a Gaussian of appropriate width.
For the modelled spectra, we simply smooth with a Gaussian of 7.9\arcsec width.  
As already seen above with the 7-9\,$\mu$m and 22-24\,$\mu$m profile comparison, the modelled spectrum
shows an excess of AIB emission relatively to the mid-IR continuum.
Furthermore, the model shows that the mid-IR continuum in the IRS
range is dominated by VSG emission. We can therefore discuss the AIBs to mid-IR continuum ratio 
in terms of PAH/VSG relative abundance.

Fig.\,\ref{fig:spectre_model_HH} also shows the case where the PAH abundance
is reduced by a factor of 2.4 with respect to that of Cirrus: the match to the data then becomes very good.
We thus conclude that the PAH abundance relative to VSGs
is 2.4 times higher in the Cirrus medium than at the peak emission position
of the Horsehead Nebula.

\subsection{Effect of differential excitation}

\begin{figure}
   \centering
     \includegraphics[width=0.47\textwidth]{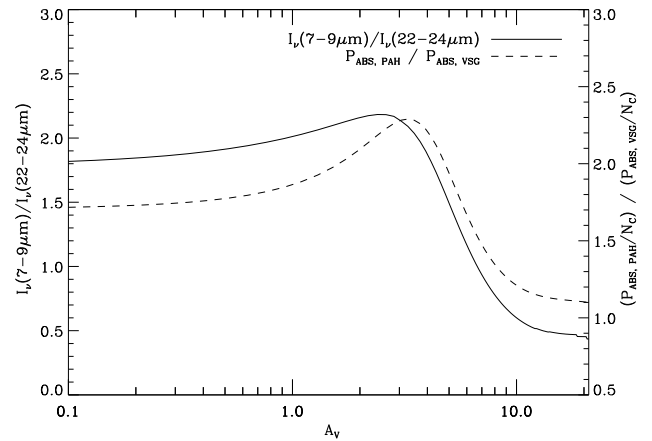}
      \caption{ The 7-9\,$\mu$m / 22-24\,$\mu$m emission ratio and the
        ratio of absorbed power per C atom by PAHs and VSGs 
        as a function of the optical depth into the
        Horsehead Nebula.
}
        \label{fig:evol_emiss_model}
\end{figure}
Here we discuss and quantify the effect on the mid-IR emission spectrum
of differential excitation of PAHs and VSGs
due to radiative transfer.
As discussed in \S\,\ref{sect:dust_model}, 
differences in the absorption cross sections
can have an impact on the relative excitation of different dust populations
depending on the spectral shape of the exciting radiation field.

Fig.\,\ref{fig:evol_emiss_model}
shows the modelled 7-9\,$\mu$m\,/\,22-24\,$\mu$m 
(i.e AIBs\,/\,mid-IR continuum)
ratio evolution as the function of the optical depth into the
cloud for the Horsehead Nebula model as describe above.
The dust properties used are those
of Cirrus. The column density (output of the model) was converted 
to $A_V$ with the \citet{Bohlin78} ratio, 
$N_H/A_V\,\sim\,1.87\,10^{21}\,cm^{-2}\,mag^{-1}$.
We see that the ratio increases by $\sim$\,25\%
at $0.1\,\la\,A_V\,\la\,3$ and then
decreases by a factor of $\sim$\,4.5 at 
$A_V\,\ga$\,10.
Fig.\,\ref{fig:evol_emiss_model}
also shows that the 7-9\,$\mu$m / 22-24\,$\mu$m ratio evolution is 
related to the evolution
of the absorbed power by PAHs and VSGs (per C atom) as a function of depth.
Note that differences between the two curves are related 
to the fact that variations of the temperature distribution
of PAHs and VSGs (which involve stochastic heating) as a function of the optical depth also
cause spectral shape evolution of the individual population 
emission spectra.

\begin{figure}
   \centering
     \includegraphics[width=0.47\textwidth]{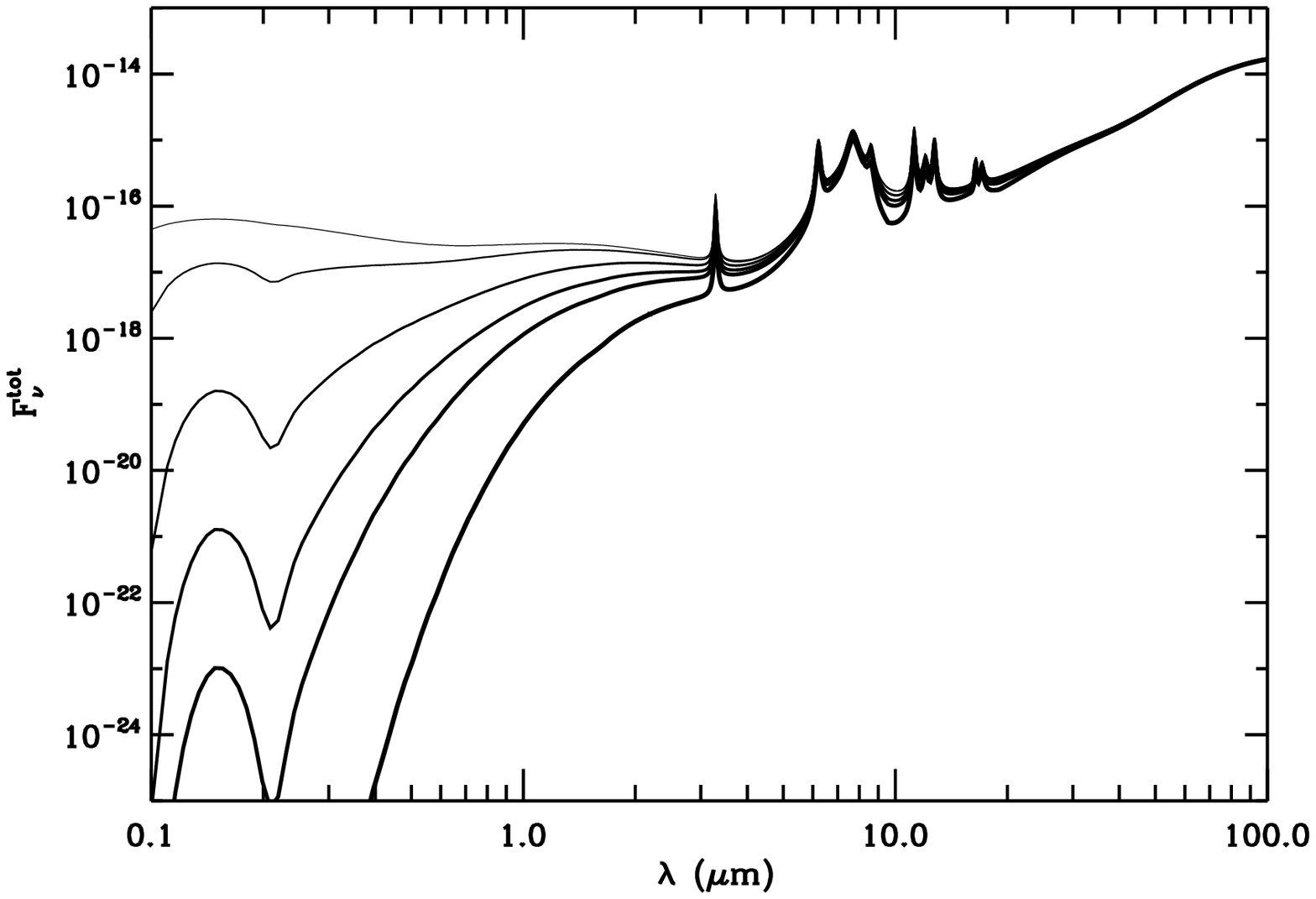} \\
     \includegraphics[width=0.47\textwidth]{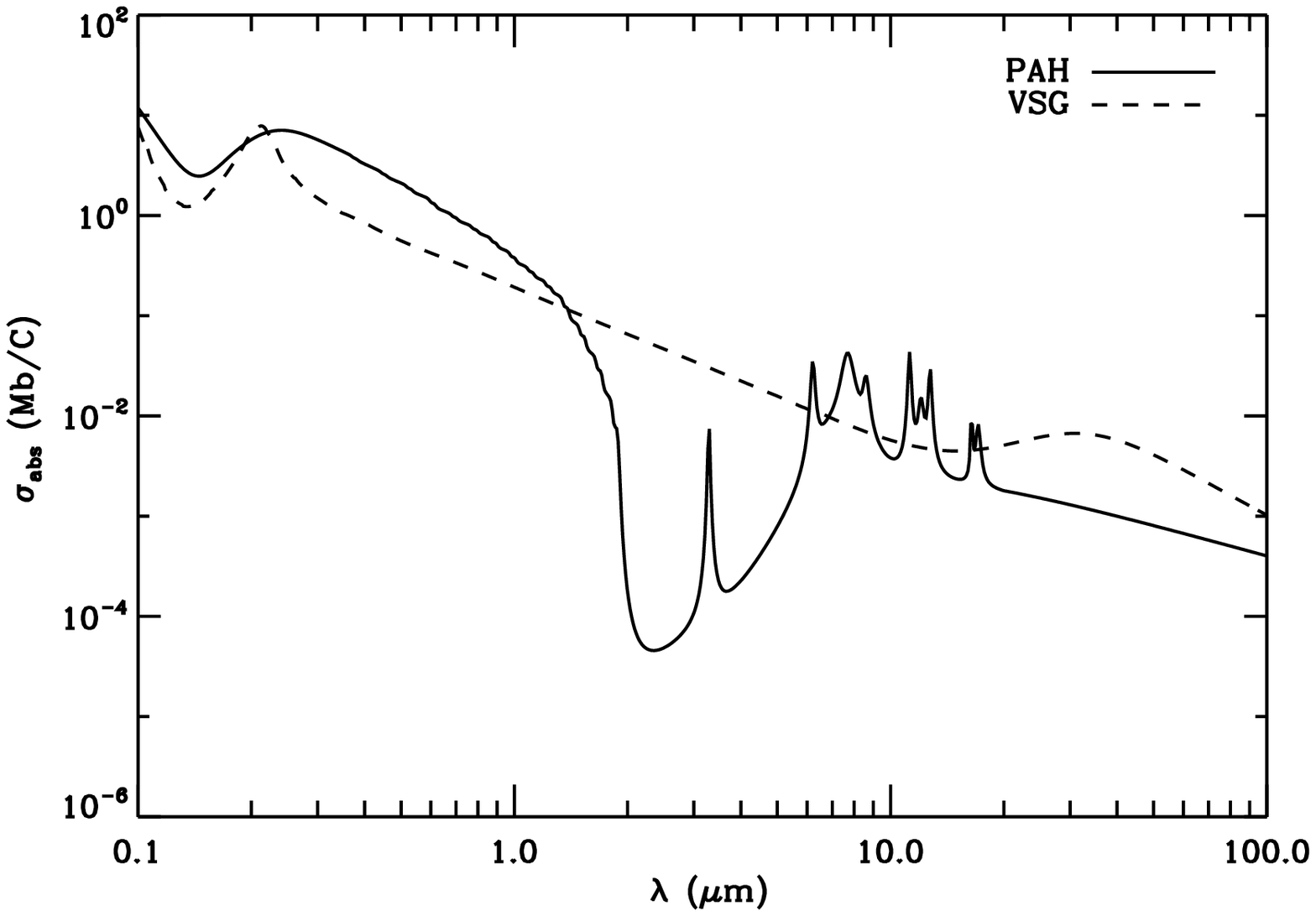}
      \caption{ {\bf Upper panel:}
        Exciting radiation field $F^{\rm{tot}}(z, \lambda)$ for $A_V$\,=\,0.1, 1, 4, 7, 10,
        20 with increasing thickness of the line for increasing
        $A_V$ and ({\bf lower panel}) the absorption cross
      sections per C atom of PAHs and VSGs of the model
      (1\,Mb\,=\,$10^{-18}\,cm^2$).
}
        \label{fig:evol_radfield}
\end{figure}

Qualitatively, this evolution of the absorbed power ratio can be understood by looking at the 
spectral evolution of the exciting radiation field $F^{\rm{tot}}(z, \lambda)$
(upper panel Fig.\,\ref{fig:evol_radfield}) and 
the cross sections of PAHs and VSGs per C atom (lower panel Fig.\,\ref{fig:evol_radfield}).
The absorption cross section of PAHs is greater than those of
VSGs for $\lambda\,\la\,$1.5\,$\mu$m
and lower for $1.5\,\la\,\lambda\,\la\,6\,\mu$m.
Hence, the reddening of the exciting radiation field
caused by extinction
successively favours PAH and VSG excitation.

We emphasize that the observed AIBs to mid-IR continuum
ratio of the Horsehead nebula cannot be explained by such differential excitation.
In fact, the 7-9\,$\mu$m / 22-24\,$\mu$m ratio is equal to 0.9
for the peak emission spectra (Fig.\,\ref{fig:spectre_model_HH}),
compared to the Cirrus value of 2.1.
In the model, the peak emission occurs at $0.2\,\la\,A_V\,\la\,0.8$
(where the profile at 7-9\,$\mu$m is equal to 2/3 of the peak value following the definition of the peak spectrum).
For this depth, the ratio does not vary by more than 13\% of the Cirrus value.
To account for the observed variations with only the radiative transfer effect, 
the mid-IR spectrum at the peak position should be emitted by dust at a depth of $A_V\,\sim\,$7, 
corresponding to a decrease of more than three orders of magnitude of the dust mid-IR emissivity 
compared to $A_V\,=\,$0.

\section{Modelling of NGC2023N}\label{sect:model_2023}

As already pointed out in section\,\ref{sect:result_obs} \citep[see also][]{abergel2002}, 
unlike for the Horsehead nebula, we spatially resolve the spectral shape evolution in NGC2023N.
For the two spectra of Fig.\,\ref{fig:LW2surLW32023}, the 
7-9\,$\mu$m / 22-24\,$\mu$m ratio goes from 1.9 
close to the star (lower spectrum of Fig.\,\ref{fig:LW2surLW32023}) 
to 0.4 for the northern position (upper spectrum of Fig.\,\ref{fig:LW2surLW32023}). 
The former must be emitted by a relatively diffuse medium
around the star just at the external edge of the dense illuminated ridge 
traced by the H$_2$ $\nu$\,=\,1-0\,S(1) emission
while the latter is emitted by denser gas at the ridge.

\subsection{The cavity spectrum}\label{sect:cavity_spectrum_2023}

\begin{figure}
   \centering
     \includegraphics[width=0.47\textwidth]{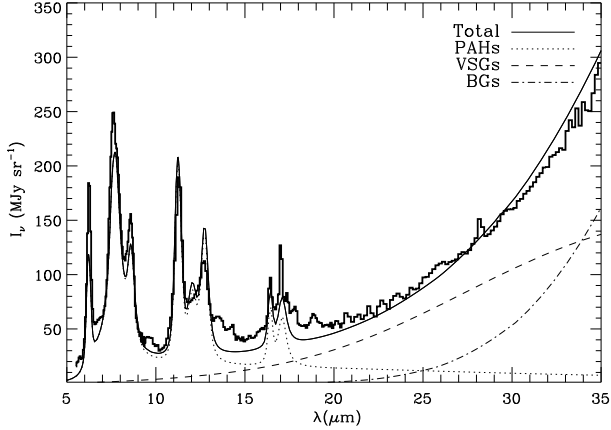}
          \caption{Comparison between the modelled and observed spectra of the
                   external diffuse part of the PDR in NGC2023N. The position of the observed spectrum is 
                   the same as the lower spectra of Fig.\,\ref{fig:LW2surLW32023}.  
                   The observed spectrum is ISOCAM-CVF at 5-14.5\,$\mu$m and IRS-LL 
                   at longer wavelength.  
                   The modelled spectrum was brought to the 
                   ISOCAM-CVF spectral resolution for $\lambda<15\,\mu$m and 
                   IRS Low-Res for $\lambda>15\,\mu$m for comparison.
                   The dust properties are those of Cirrus.}
         \label{fig:spectre_2023_cavite}
\end{figure}

It is interesting to note that the 7-9\,$\mu$m / 22-24\,$\mu$m
ratio of the spectrum from the diffuse illuminated part of the PDR ($\sim$1.9) is close to the Cirrus value ($\sim$2.1).
As shown in Fig.\,\ref{fig:spectre_2023_cavite}, the
observed spectra can be reproduced by the model by using Cirrus 
dust properties.
The modelled spectrum 
was obtained by considering a star of 
$T_{eff}$\,=\,23700\,K and $R_\star$\,=\,6\,$R_\odot$ (for B1.5V
star) located at 0.3\,pc which corresponds to the $\sim$150\arcsec\, between 
the star and the location of the observed spectrum for a distance of 400\,pc.
The slope of the continuum emission is reproduced by applying an extinction 
of $A_V\,=$\,1.25 between the star and the emitting dust. Less extinction would result in hotter BGs 
and in a too steep slope of the continuum while more extinction would result in colder BGs and a slope not steep enough. 
That extinction is consistent with the presence of matter in the cavity \citep{witt84, burgh2002}. Moreover,
this value of $A_V\sim$\,1.25 caused by the cavity matter is of the same order as
that reported between the Earth and the star by \citet{burgh2002}, which is $A_V\,\sim$\,1.4. 
It is not solely the extinction that allows us to adjust
the BG emission. 
Considering the error bars on the distance estimate of NGC2023,  
an underestimation of this distance would cause
an underestimation of the applied dilution of the star radiation field in the model.
The absolute intensity of the modelled spectrum matches that of the observed one for a column density 
of $N_H\,=\,4.6\,10^{21}\,cm^{-2}$ (in the line of sight).

As suggested in section\,\ref{sect:result_obs}, we see that the non-linear rise of
the mid-IR continuum can be explained by a significant emission
of BGs at $\lambda\,\ga\,25\,\mu$m while according to the model
the continuum emitted by the VSGs is mostly linear. 
Note that the intensity of the exciting radiation field
after dilution and extinction is $\chi\sim$120 (in unit of the
\citet{Habing68} field between 6 and 13.6\,eV) which is enough for
the BGs to emit noticeably in the IRS spectral range.

BGs do not emit significantly in the 22-24\,$\mu$m range at this position and will be colder 
further from the star which make them emit at longer wavelengths.
The decrease by a factor of $\sim$\,5 of the  7-9\,$\mu$m / 22-24\,$\mu$m ratio
may then be related to the evolution of the PAH/VSG relative emission,
as in the Horsehead case.
This is coherent with the linear shape of the 
mid-IR continuum for the observed spectrum deep into the dense ridge
(Upper spectrum of Fig.\,\ref{fig:LW2surLW32023})

AIB spectrum is also well reproduced by the Cirrus
PAH properties which suggest similar properties of these emitters
in the diffuse medium around the star and in Cirrus.

\subsection{The spectral shape evolution}

As seen in Fig.\ref{fig:LW2surLW32023}, the spectral shape evolves
in the dense ridge traced by the H$_2$ $\nu$\,=\,1-0\,S(1) emission
located just after the location of the studied cavity spectrum.
As in the case of the Horsehead, we can model the evolution 
of the 7-9\,$\mu$m\,/\,22-24\,$\mu$m emission ratio as a function of the optical depth
into the dense illuminated ridge in order
to quantify the radiative transfer effects.
Unlike the Horsehead nebula, the density profile is not constrained.
However, as reported 
by \citet{field98} (see references therein), the density must range between
$10^4$ and $10^5\,cm^{-3}$ in the dense illuminated ridge as shown through the use
of steady-state PDR models to study
the brighter part of NGC2023. 
The simplest model for the dense ridge 
is then a density wall illuminated   
by the B1.5V star located at 0.3\,pc.
The radiation field of the star that illuminates the 
ridge must be extinguished with 
$A_V\,=$\,1.25 to be consistent with the value of the extinction at the
location of the studied cavity spectrum, which is the position where
the dense ridge arises. Note that this extinction corresponds to a constant density of $n_H\,=\,2550\,cm^{-3}$
for the cavity.
\begin{figure}
   \centering
    \includegraphics[width=0.47\textwidth]{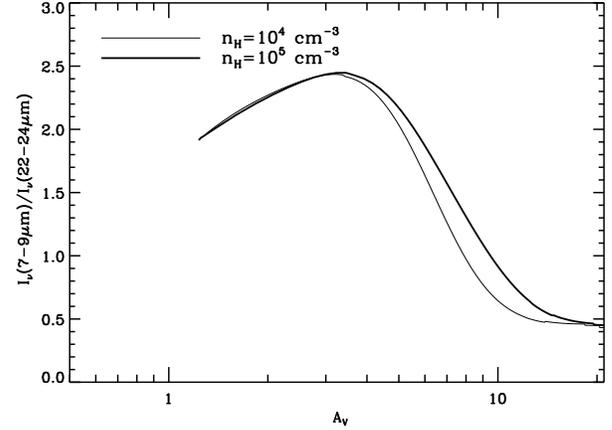}
          \caption{7-9\,$\mu$m / 22-24\,$\mu$m emission ratio as a function of the depth in the dense illuminated ridge of NGC2023N.}
         \label{fig:evol_ratio_7p7_cont_2023}
\end{figure}

Fig.\ref{fig:evol_ratio_7p7_cont_2023} shows the 
result obtained for the 7-9\,$\mu$m / 22-24\,$\mu$m emission ratio as a function of the optical depth
in the dense ridge for
densities of $10^4$ and $10^5\,cm^{-3}$.
We can see that results for the two different densities are similar and also are
similar to the result obtained in the Horsehead case.
Hence, radiative transfer effects cannot
explain the spectral variation in NGC2023N since 
a depth of $A_V\,\ga\,$12 is required to explain the  
value of 0.4 for the 7-9\,$\mu$m / 22-24\,$\mu$m
ratio observed in the deep/dense part of the ridge.
Such an extinction would cause a decrease of more than 3 orders of magnitude
of the the 22-24\,$\mu$m dust emissivity compared to the edge of the ridge ($A_V\,=$\,1.25)
while the observed intensity increases from $\sim$70\,MJy\,sr$^{-1}$ to $\sim$100\,MJy\,sr$^{-1}$
between the cavity spectrum and the deep/dense ridge spectrum.
We can conclude that
the PAH/VSG relative abundance increases by up to a factor of $\sim\,$5 from the deep/dense part 
to the diffuse/illuminated part of NGC2023N. 
Moreover, dust properties
in the diffuse medium surrounding the star in NGC2023
must be the same 
as in the diffuse high galactic latitude medium (Cirrus).

\section{Conclusion}\label{sect:conclusion}

We presented mid-IR spectral imaging observations of the Horsehead nebula
and NGC2023N 
obtained with the infrared spectrograph on board the Spitzer Space Telescope.
These observations allow us to confirm 
the AIBs / mid-IR continuum evolution at dense illuminated 
ridges already observed with the Infrared Space Observatory \citep[e.g.][]{abergel2002, rapacioli2005}.

We developed a new dust emission model based on the
\citet{desert90} model. 
This model successfully reproduces the emission and
the extinction curve of the diffuse interstellar medium
at high galactic latitude (i.e. the Cirrus).
The Cirrus dust properties are then  
used as a reference to be compared to dust properties
in the two studied PDRs.
This dust emissivity model is coupled to a radiative 
transfer model in order to properly take into account 
excitation effects on the mid-IR spectral shape evolution.

For the PDRs studied here, we conclude that
excitation effects cannot account for the observed 
AIBs / mid-IR continuum evolution.
We interpret this evolution in term of PAH/VSG relative abundance variation since 
these two species
dominate the 5-35\,$\mu$m spectrum of the two studied PDRs except at
$\lambda\,\ga25\,\mu$m for positions
close to the star (in the cavity) in NGC2023 where the BG continuum is significant.

In the Horsehead Nebula case, we do not spatially resolve any spectral variation.
We conclude from the study of
7-9\,$\mu$m (AIBs) and 22-24\,$\mu$m (mid-IR continuum) emission profiles
and of the spectrum at the emission peak position that the PAH/VSG relative
abundance is 2.4 times lower than in the Cirrus.
In NGC2023N, we resolve the spectral variation.
We modelled the spectrum of the diffuse medium around the star
for a position at the external edge of the dense illuminated ridge traced by H$_2$ $\nu$\,=\,1-0\,S(1).
This spectrum can be reproduced by using Cirrus dust properties.
We attributed the decrease of a factor of $\sim$5 of the 7-9\,$\mu$m\,/\,22-24\,$\mu$m
(AIBs / mid-IR continuum)
from the diffuse illuminated part to the deep/dense
part of the PDR to a decrease of the PAH/VSG relative abundance
to $\sim$1/5 of the Cirrus value.
It then seems that dust properties evolve fully "from dense to diffuse" properties
within the small spatial scale of the dense illuminated ridge.
Note that the obtained PAH\,/\,VSG relative abundance
for the "unresolved" Horsehead is almost the median between the diffuse 
and dense part of NGC2023N, indicating the mean properties of the observed
very small particles in the Horsehead.

An increase of the PAH/VSG relative abundance from the dense/deep to
the diffuse illuminated part of the PDR is fully consistent with a
scenario of photoevaporation PAH clusters (i.e, of the VSGs)
\citep[e.g.][]{cesarsky2000,rapacioli2005, berne2007}.
Other physical processes could explain this variation,
such as (i) the scenario evoked in \citet{jones2005} of the evolution of
aliphatic hydrocarbons (which do not emit AIBs) which are present in
the dense media \citep[e.g.][]{jones90, jones90b, dartois2007} into
aromatic hydrocarbons under the effect of the UV radiation field
\citep[e.g.][]{dartois2005, jones90}, (ii) the release
("decoagulation") of the VSGs and PAHs from the dust aggregates
(present in dense clouds) at different depths into the PDRs or (iii)
a size segregation effect due to the grain dynamics induced by the
anisotropic radiation field, as already mentionned for NGC2023 by
\citet{abergel2002}.  All these processes could be at work with
different efficiencies depending on the depth into the cloud and/or on
the size and nature of dust.

This strong evolution of the PAH relative abundance 
between the dense and diffuse medium in PDRs could be a clue for the interpretation of
the observed (PAH 8\,$\mu$m)/24\,$\mu$m and (PAH 8\,$\mu$m)/160\,$\mu$m variations 
in the SINGS galaxies \citep[e.g.][]{bendo2008}.
Such a PAH relative abundance evolution in the bright IR emitters
that are PDRs will have an impact on the
estimation of the star formation activity of galaxies using PAH emission.

\begin{acknowledgements}

The authors wish to acknowledge all the members
of the SPECPDR team for their contribution to the success of the proposal.
We would like to thank the referee for a constructive report.
This work is based on Spitzer and ISO observations.
ISO was an ESA project with instruments funded by ESA Member States
and with the participation of ISAS and NASA.
Spitzer is operated by the Jet Propulsion Laboratory,
California Institute of Technology under a contract with NASA.

\end{acknowledgements}

% for the bibliography, at the end
\bibliographystyle{aa} % style aa.bst
\bibliography{09850}

\end{document}